\documentclass[leqno]{article}
\usepackage[pdfborder=0 0 0,colorlinks=true,linkcolor=blue]{hyperref}
\usepackage{graphicx}
\DeclareGraphicsExtensions{.eps}
\usepackage[letterpaper,left=3cm,top=3.5cm,right=3.5cm,bottom=6cm]{geometry}
\usepackage{color}
\usepackage{amsmath}
\usepackage{amssymb}
\usepackage{pdfpages}
\usepackage{bbold}
\usepackage[utf8x]{inputenc}
\usepackage{cancel}
\usepackage[capitalise]{cleveref}
\newcommand{\eps}{\varepsilon}

\DeclareMathOperator{\arccot}{arccot}

\begin{document}
\begin{center}

{\Large Effects of  source and loss terms on 
the wave-pinning description of cell polarisation.\\}

\vspace{0.3cm}

{\small Nicolas Verschueren\footnote{Department of Engineering
    Mathematics, University of Bristol, Queen's Building. University
    Walk, Bristol BS8 1TR, United Kingdom
    (n.verschuerenvanrees@bristol.ac.uk)} and Alan Champneys\footnote{
    Department of Engineering Mathematics, University of Bristol,
    Queen's Building. University Walk, Bristol BS8 1TR, United Kingdom
    (A.R.Champneys@bristol.ac.uk)}.}
\end{center}
\hrule
\begin{center}
\textbf{Abstract}\\
A system of two Schnakenberg-like reaction-diffusion equations is
investigated analytically and numerically. The system has previously
been used as a minimal model for concentrations of GTPases involved in
the process of cell polarisation. Source and loss terms are added,
breaking the mass conservation, which was shown previously to be
responsible for the generation of stable fronts via a so-called
wave-pinning mechanism.  The extended model gives rise to a unique
homogeneous equilibrium in the parameter region of interest, which
loses stability via a pattern formation, or Turing bifurcation. 
The bistable character of the reaction terms
ensures that this bifurcation is subcrtical for sufficiently small
values of the driving parameter multiplying the nonlinear
kinetics. This subcriticality leads to the onset of a multitude of
localised solutions, through the homoclinic snaking mechanism. As the
driving parameter is further decreased, the multitude of solutions
transforms into a single pulse through a Belyakov-Devaney transition
in which there is the loss of a precursive pattern. An asymptotic
analysis is used to probe the conservative limit in which the source
and loss terms vanish.  Matched asymptotic analysis shows that on an
infinite domain the pulse solution transitions into a pair of fronts,
with an additional weak quadratic core and exponential tails. On a
finite domain, the core and tails disappear, leading to the mere
wave-pinning front and its mirror image.
\end{center}
\begin{flushleft}

\textbf{Key words:} Reaction-diffusion systems. Cell polarisation. G-proteins.\\
\textbf{AMS subject classifications.} 35B25, 35B32, 35K57, 34B07\\
\end{flushleft}

\section{Introduction}

Eukaryotic cell polarisation is the process by which a cell forms two
distinct spatial domains a ``front-end'' and a ``back-end'', defining
 a polarisation axis.  This process is the first step in many
vital cellular processes such as cell differentiation, wound healing, cell
motility and organelle organisation.  Roughly speaking,
cell polarisation can be described as the symmetry-breaking presented
by the spatial concentration of certain proteins and lipids inside the
cell \cite{revjilkine}. Polarisation may be caused spontaneously or by some
external trigger or stimulus,  here we shall consider the polarisation
induced by an external stimulus, acting on the  spatially
heterogeneous concentration of certain GTPases, also known as G-proteins or
Rho's. See \cref{sec:wpm} below for background information.  

The spatio-temporal dynamics of  Rho-GTPases within a single cell has
been described using reaction-diffusion models; see \cite{revleah, revjilkine, rev3}. 
 Minimal models consider just one
GTPase, present in both active and inactive forms.  Examples of such
models were proposed by Meinhardt \cite{cones} and, with the explicit
introduction of mass conservation, by Otsuji {\em et al.}
\cite{otsuji}. These works presume that the fundamental mechanism for
spatial symmetry breaking is the so-called diffusion driven or Turing
instability, see e.g.~\cite{Murray2}. This mechanism is appealing,
because there is a natural discrepancy between the diffusion rates of
the free, inactive GTPases and their membrane-bound active
counterparts. However, the Turing mechanism is known to lead to 
patterns characterised by a specific spatial wavelength rather
than a single front or pulse ---
see.~\cref{fig:cellmodelling}(a,b) below. 
(Note though that the assumption in these works
is that of supercrtical Turing bifurcations; it is argued
in \cite{Victor3} that
subcritical Turing bifurcations
naturally lead to spatially localised states through
the so-called homoclinic snaking mechanism \cite{snakebeck,alansnake}).
Additionally, the timescale required for the break of
symmetry in the Turing instability was found not to match those observed
experimentally; see \cite{revjilkine} for details.

In order to overcome these difficulties, Mori {\em et al.} \cite{Morior},
proposed a rather different mechanism, called \emph{wave pinning}, in which
a single front is set up between two different asymptotic levels
of the active G-protein.  The pinning mechanism is different from
that of scalar reaction-diffusion equations and relies on the resting
position of a moving front being set by the overall mass conservation of
active and inactive species (see \cref{sec:energy} below 
for a concise argument). 
Once the front is
established, the polarity can be inverted (specular reflection of the
front) through new stimuli, a feature which is observed in the
experiments. This minimal model has also served as the basis for more
complex models in cell motility (see \cite{marvercom} and references
therein).

The minimal models describing the dynamics of a single GTPase are intended
to be prototype models. However they have been successfully fit
to experimental data \cite{warwick}. The simplicity of these
models allows analytic calculations which enable us to uncover the
essential mechanism responsible for cell polarisation. In \cite{Mori}
a simple bifurcation analysis of the wave pinning model is
carried out. In the same spirit, in \cite{weaklyrubinstein} linear and
weakly non-linear analysis of the Otsuji model is performed. Also 
\cite{njpger} identify a cusp
bifurcation as being the organising centre responsible
for setting up the bistable kinetic profile responsible for the cell
polarisation. 

The aim of this paper is to consider the effect of
source and loss terms on the wave-pinning model. Specifically we
are motivated by the related Schnakenberg-like  
model proposed by Payne and Grierson
\cite{Payne} and further studied in \cite{Victor1,Victor3} for
the formation of a single localised patch of active 
Rho's in {\em Arabidopsis} root hair cells. That model has striking 
similarities to the wave-pinning model, with the main difference being the
presence of source and loss terms which were argued to represent nuclear
control and secondary growth initiation processes respectively, which
occur on a similar timescale to the the patch formation. 

Specifically, in what follows, we shall study the dimensionless system of
equations 
\begin{subequations}
\label{dimle}
\begin{align}
\label{dimlea}
\partial_t u &=\delta\partial_{xx}u +[F(u,v)-\varepsilon \theta u],\\
\label{dimleb}
\partial_t v &=\partial_{xx}v - [F(u,v)-\varepsilon \alpha],\qquad x\in 
\left[ -L,L \right], \quad \partial_x (u,v) 
(\pm L)=0 , \\
\label{dimlec}
\text{where} & \quad F(u,v) =\gamma \frac{u^2v}{1+u^2}-\eta u+v. \qquad
\end{align}
\end{subequations}
Here $2 L\gg 1 $ is a large (possibly infinite) domain length, $u(x,t)$ and  $v(x,t)$ represent the concentrations of active and 
inactive species
respectively and $\delta \ll 1$ is the ratio of their diffusion rates.  
The function $F(u,v)$ represents the local kinetics of the activation
step parametrised by $\mathcal{O}(1)$ parameters $\eta$, $\gamma$. 
The specific form of $F$ is not important and indeed we shall also consider
a simpler form in \cref{sec:energy} in order to make explicit
illustrative calculations. The parameters $\alpha$ and $\theta$ represent
the strength of a constant source of inactive form and linear loss 
of active form respectively, with $\eps>0$ representing the relative
importance of these effects compared to the other dynamics. In particular
we shall be interested in both the cases $\eps =1$ and $\eps \to 0$; 
the latter case leading precisely to the wave pinning model.

The rest of the paper is organised as follows. 
 \Cref{sec:wpm} contains a brief review of the biology of cell
polarisation and of the minimal reaction-diffusion models that have been 
proposed to describe it. Then, \cref{sec:energy} contains a new, 
simplified analysis of the wave-pinning mechanism. 
 \Cref{sec:resultados} contains analytical, simulation and
numerical continuation results on the existence of stable localised
states of the model \cref{dimle} for $\eps=1$, while taking $\delta$ and 
$\gamma$ as bifurcation parameters and prototypical values of the other 
constants. Parameter regions
are identified in which 
homogeneous states, periodic states, localised patterns, or isolated pulses
may be observed. Then, the key question is addressed in \cref{sec:consv} 
of how this structure composed by different states collapses to the wave-pinned fronts
solutions in the mass-conservation limit $\eps \to 0$. First, numerical results
show how the pulse solutions transform into front 
and back pairs with non-vanishing core and tails. These results
suggest distinguished scalings that leads to a   
multiscale asymptotic analysis. This analysis also explains the 
key differences observed on a finite rather than infinite domain. 
Finally, \cref{sec:concl} draws conclusions and discusses
potential implications of our findings to the biology of cell polarisation
and more generally to pattern formation theory.

\section{Cell polarisation models}
\label{sec:wpm}

 Eukaryotic cells can respond to gradients caused by 
small differences in concentration of exogenous or internal chemical signals. 
The phenomenon of cell polarisation induced by such stimuli has been observed experimentally in several cell types such as: budding yeast, \emph{Dictyostelium discoideum} and Mammalian cells (white blood cells) \cite{smithyeast}. 
From these experiments, many  factors involved in the cell
polarisation have been identified,  a few being proteins like small
GTPases (Cdc42, RaC and Rho), PI membrane lipids (PIP, PIP$_2$
 and PIP$_3$) and Arp 2/3 (in active cytosol form).  The interaction and
spatial concentration of these factors depend on the
 cell-type, position and  cellular state.  Moreover, the  molecular networks
responsible for cell motility or chemotaxis are complex and depend on
the cell-type. However, the basic mechanisms seemed to be preserved
across all eukaryotic cells even though not all these factors
are present in every cell type. 

\begin{figure}
\begin{center}
\includegraphics[width=12cm]{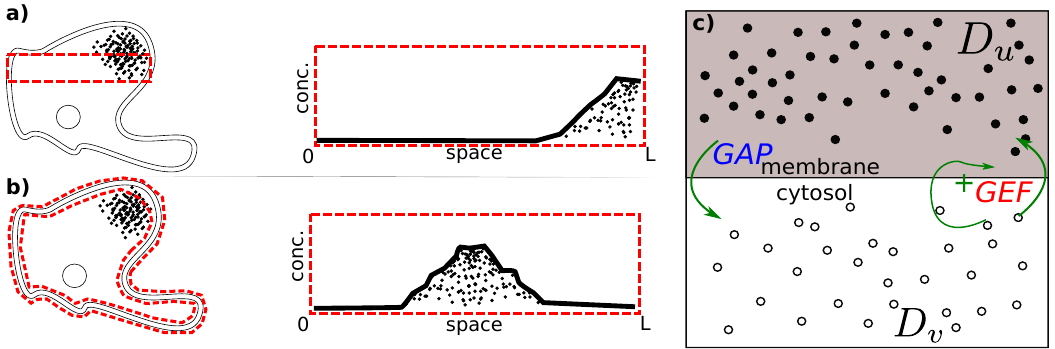}
\end{center}
\caption{Schematic description of the mathematical modelling. 
(a,b) 
Representation of the spatial domain as either  a typical radial ((a), left)) 
or circumferential ((b), left) slice of the cell. 
Consequently polarisation corresponds to either a front solution
 ((a), right) or a localised solution ((b), right). (c) 
The modelled protein exists in an active form in the membrane, with diffusion
constant $D_u$ and the
inactive form is in the cytosol (with active form $D_v$).   
The protein is deactivated by GAPs and activated through GEFs, the latter
of which is assumed to be autocatalytic.} 
 \label{fig:cellmodelling}
\end{figure}

Consequently, in order to study cell polarisation as a phenomenon
independent of cell-type, a minimal approach is often used, where the
features that are common to all cell-types studied are captured. The
most relevant factors appear to be the small GTPases, known
collectively as Rho proteins. These proteins are present in the
cytosol (in inactive GDP-bound form) and the membrane (in active
GTP-bound form). There is a constant exchange between the active and
inactive forms. The active form can be deactivated via GTPase activating
proteins (GAPs) and the inactive forms can be activated by Guanine
exchange factors (GEFs).  GEFs are thought to be responsible for the
observed positive feedback, or autocatalysis of the activation step. 
Moreover, experimental observations have shown that there is a big difference
between the diffusion ratios of G-proteins in the membrane and the
cytosol (\cref{fig:cellmodelling}(c)). Finally, within the timescale
of polarisation (minutes), the total amount of the protein
(considering both active and inactive form) is often taken to be
constant \cite{sandrine}.

The spatio-temporal dynamics of the concentration of G-proteins in the
cell (whether in active or inactive form) are determined by two
processes: \emph{diffusion} and \emph{reaction}. The diffusion is
responsible for the spatial dependence. On the other hand, GAPs and
GEFs induce local reactions that are modelled typically by the
non-linear interplay between components. 

In \cite{Morior}, the wave pinning model ( expression \cref{dimle}
when $\varepsilon=0$)
was introduced as a minimal reaction-diffusion 
description for the cell polarisation problem. 
In their derivation, the unidimensional domain is obtained from 
taking a typical radial slice through the cell. 
Throughout the spatial domain, both 
\emph{cytosol} and \emph{membrane} are present. 
In the \cref{fig:cellmodelling} 
a sketch of the domain is depicted. In this scenario, 
the cell polarity is characterised by increased concentration 
of active G-proteins 
(black dots in the \cref{fig:cellmodelling}(c)). In the continuum
limit, such a state would correspond to a front connecting low to high
concentrations of the 
active form (\cref{fig:cellmodelling}(b), right).
Thus, the wave-pinning model considers the spatio-temporal dynamics of 
the concentration of a single G-protein, existing in active and inactive form in
the same long spatial interval.

To explain where the model comes from, let $\hat{u}(\hat{x},t^*)$ and
$\hat{v}(\hat{x},t^*)$ represent the concentrations of the active and
inactive forms.  Experimental findings show that there is at least a
ten-fold difference between the diffusion rates of the active and inactive
forms; in dimensional co-ordinates $D_u\ll D_v$. The conservation of
total concentration suggests both the use of a non-flux boundary
conditions and the same kinetic function $\hat{F}$ for creation of
active and destruction of inactive forms.  The function $\hat{F}$
accounts for both the activation and deactivation steps, mediated by
the GEF and GAP respectively. Let $k_0$ and $\bar{\delta}$ represent
the basal rates of each process.  The positive feedback of the
activation can be modelled by a \emph{Hill function}, the simplest
form of which should have coefficient 2, in order to enable the
appropriate symmetry consideration. Letting the Hill parameters be
given by $\gamma$ and $K^2$, we have the simplest form
$$
\hat{F}(\hat u,\hat v)=\gamma \frac{\hat{u}^2 \hat{v}}{K^2+\hat{u}^2}-\bar{\delta} \hat{u}+k_0 \hat{v}.
$$
Actually, as discussed and explored in \cite{Mori}, any reasonable 
function can be used, as long as it satisfies certain
non-degeneracy assumptions.  Given $\hat{F}$, the full model is
\begin{subequations}
\label{original}
\begin{align}
\label{eqa}
\partial_{t^*} \hat u &= D_u \partial_{\bar{x}\bar{x}} \hat u +\hat F(\hat u,\hat v),\\
\label{eqb}
\partial_{t^*} \hat v &= D_v \partial_{\bar{x}\bar{x}}  \hat v-\hat F(\hat u,\hat v), \qquad \bar{x} \in [-L,L], \quad \partial_{\hat{x}} (u,v) (\pm L) =0,  \\
\label{eqc}
T &= \int_{-L}^L (\hat u+\hat v) d\bar{x}. 
\end{align}
\end{subequations}
Here $T$ is proportional to the total mass of G-proteins in the domain. 
Note, from the form of these equations and the no-flux boundary conditions 
that $\partial_{t} T=0$, so that the total mass is conserved. 

\begin{figure}
\begin{center}
\includegraphics[width=7.3cm]{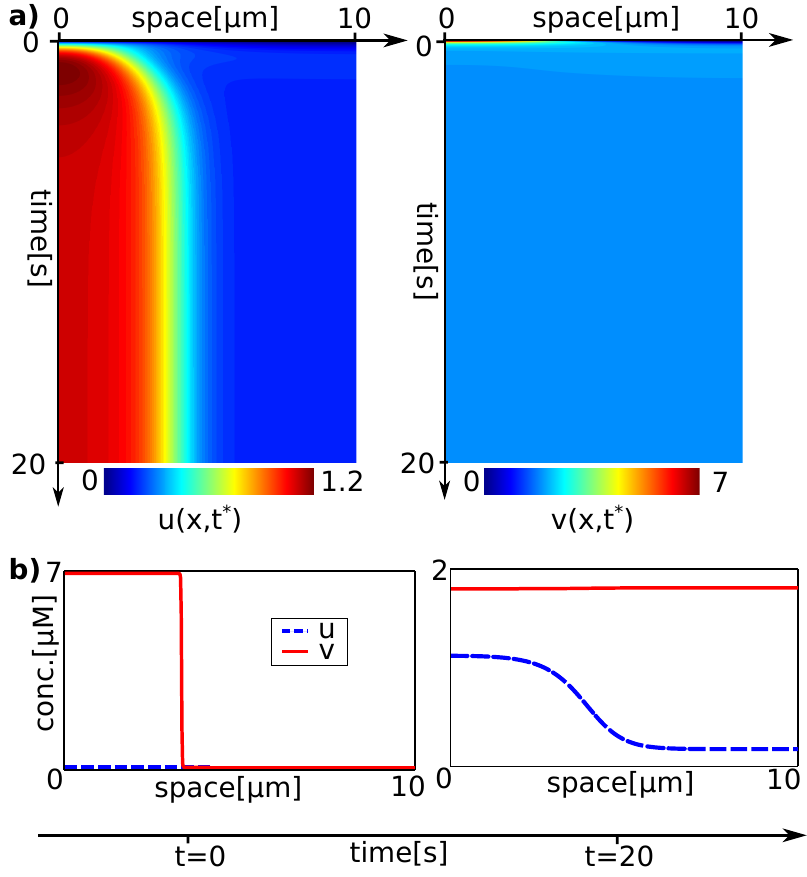}
\end{center}
\caption{Numerical observation of the wave-pinning phenomenon in model \cref{original}; (a) spatio-temporal plots, (b) initial and
final solutions. See text for details. 
Parameter values are $2L=10[\mu m], \delta=1[s^{-1}], \gamma=1[s^{-1}],K=1[\mu M],k_0=0.067[s^{-1}], D_u=0.1[\mu m^2 s^{-1}],D_v=10[\mu m^2 s^{-1}].$}
 \label{fig:figura1}
\end{figure}

 \Cref{fig:figura1} shows simulation results for the 
model at typical parameter values. 
Initially (b, left), an asymmetrical stimulus 
is applied for the inactive form $v$ (continuous red line) 
while the active form $u$ remains homogeneous (blue dashed line). 
Consequently, the active form
$u(x,t)$ develops a front solution that propagates as the inactive
form tends to a roughly homogeneous equilibrium (see the spatio-temporal
plots in (a)). After a characteristic
time, a steady front is established in the active form (cf.~the dashed
blue line in (b), right).

An in depth treatment of
the wave-pinning model and its various extensions 
can be found in \cite{Morior,Mori}. An energetic explanation
of the pinning phenomenon is given in \cref{sec:energy}.

According to the theory developed in \cite{Pomeau}, a front connecting
two given homogeneous equilibria will be stable only at one point in
the parameter space called the \emph{Maxwell point} 
(see \cref{sec:energy} for more details). 
In the case of model \cref{original}, it possess a continuum of
homogeneous equilibria given by the curve $F(u,v)=0$ which is the nullcine
of both the $u$ and $v$ spatially-independent systems
(cf.~\cref{fig:coor}(a)). This degeneracy provides an extra
degree of freedom in the system which turns the \emph{Maxwell Point}
into an interval of parameter values. 

When an asymmetrical stimulus is applied to $v$ and a monotonic front 
arises, then the homogeneous equilibria either side of
the front are free to move
along the nullcline $F=0$, attaining a stable stationary 
front whose
homogeneous equilibria lie in the \emph{Maxwell Region}, a subset of
the \emph{bistability region}. The precise choice of which among the family
of possible front solutions is chosen is determined by the initial
total mass $T$.

In comparison with other proposed models, the 
wave pinning mechanism seems to account for several typical features
of experimental observations (see \cite{revjilkine} for a review). For instance: the polarisation time in the model ($\sim 20s$) is in good agreement with experiments, the homogeneous states are stable and the front can be reversed through the introduction of new stimuli. 
Due to its popularity, the wave pinning model has also served as a basis 
for further investigation of cell polarisation phenomena \cite{Mare,njpger}.

It is useful to nondimensionalise the wave-pinning model by introducing the dimensionless quantities:
\begin{align*}
 t &=k_0 t^*, \; x=\bar{x}\sqrt{\frac{L^2k_0}{D_v}}=\bar{x}\mathcal{L},\;  \delta=\frac{D_u}{D_v},\\
 \gamma &=\frac{\bar{\gamma}}{k_0},\; \eta=\frac{\delta}{k_0},\; \hat u=K u,\; \hat v=K v,
\end{align*}
 the model \cref{original} takes the form
\begin{subequations}
\label{dimlex}
\begin{align}
\label{dimleaz}
\partial_t u &=\delta\partial_{xx}u +F(u,v),\\
\label{dimlebz}
\partial_t v &=\partial_{xx}v - F(u,v) \quad x\in \left[-\mathcal{L},\mathcal{L}\right],\\
\label{dimlecz}
\text{where}& \quad F(u,v) =\gamma \frac{u^2v}{1+u^2}-\eta u+v.
\end{align}
\end{subequations}

We can break the mass conservation law by adding generic source and loss terms 
(as in \cite{Victor2}). The extra control parameter $\varepsilon$ is introduced in order to investigate the r\^{o}le played by the new terms. The new extended model therefore takes the dimensionless form \cref{dimle}

\section{Energetic description of the wave-pinning phenomenon} 
\label{sec:energy}

The above-described phenomenon of wave pinning can be understood in 
terms of Maxwell-point theory as follows. Consider the 
dimensionless wave-pinning model \cref{dimlex} and introduce 
the new variables 
$$
S(x)=\delta u(x)+v(x), \quad R(x)=\delta u(x)-v(x). 
$$
The spatial system then takes the form
\begin{subequations}
\label{wpnv}
\begin{align}
\label{wpnv:R}
\frac{d^2 R}{dx^2}&=0, \\
\label{wpnv:S}
\frac{d^2 S}{dx^2} &= -2 F(S,R_*) :=-\frac{d V}{d S}(S,R_*). 
\end{align}
\end{subequations}

\begin{figure}
\begin{center}
\includegraphics[width=15cm]{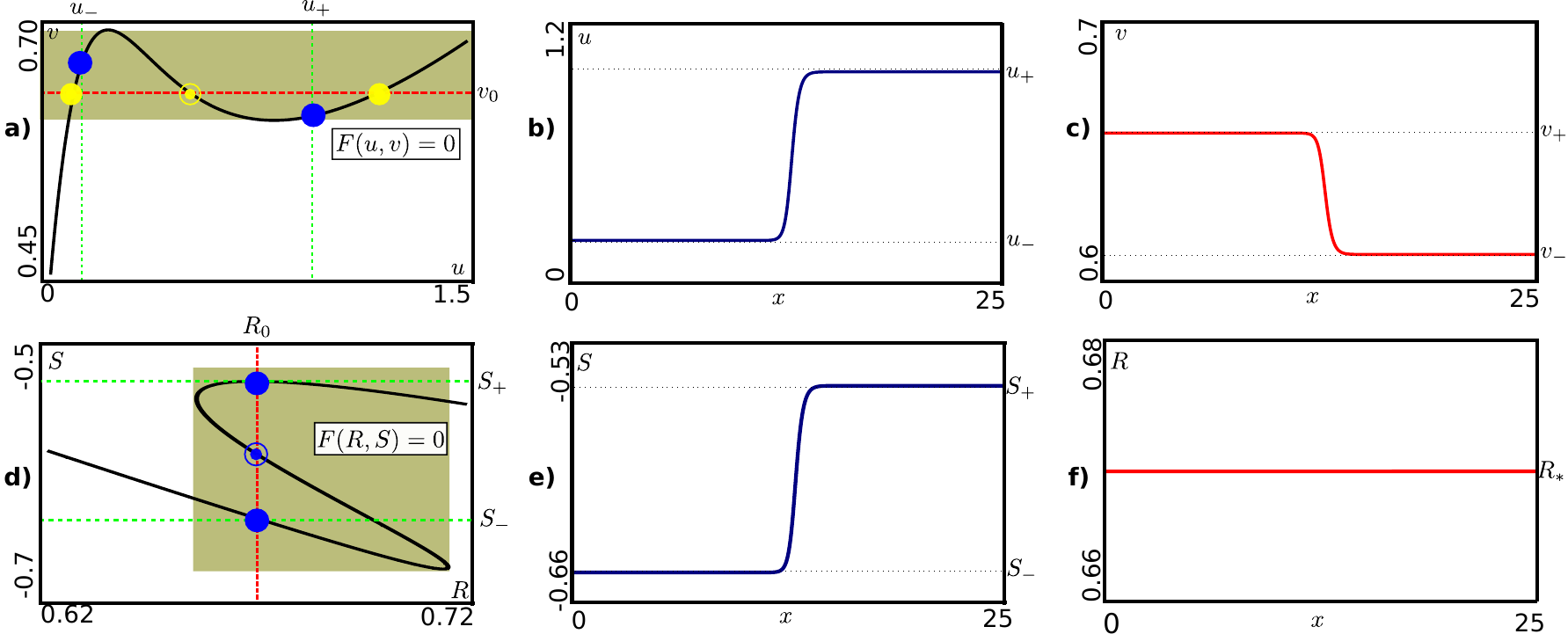}
\end{center}
\caption{Illustrating the wave pinning phenomenon in two different 
coordinates systems, see
text for details. Parameters values used are
$\delta=0.06$, $\gamma=15$, $\eta=5.2$, $2L=25$. } 
\label{fig:coor}
\end{figure}

Owing to the boundary conditions, the solution for the $R$ variable is
a constant, $R=R_*$ say. Hence 
the four-dimensional spatial dynamical system 
is reduced to a two-dimensional one for the variable $S$, albeit with
an unknown parameter $R_*$. This new formulation enables a simpler 
description of the wave-pinning phenomenon. In \cref{fig:coor},
 the panels (a) and (d) depict the nullclines in both the $(u,v)$ and $(R,S)$ coordinate
systems, with the green areas demarcating the bistability region. 
Full and empty circles correspond to homogeneous equilibria that
would be stable or unstable respectively in the absence of diffusion. 
Considering the front solution in the case
$\delta \ll 1$, the changes in $v$ along the front are negligible
compared with the changes in $u$. Therefore, the inactive form 
is approximately constant ($v(x,t) \approx v_0$, note the small
vertical scale in panel (c)). 
In \cref{fig:coor}(a), the line solution for $v_0$ is depicted by a 
horizontal red dashed line.  The yellow
points correspond to the possible homogeneous equilibrium values
for $u$ when $v=v_0$. The blue points mark the values for
$(u_-,v_+)$ and $(u_+,v_-)$ for the front solution. The components $u$
and $v$ are illustrated in (b) and (c). The corresponding
pictures for the $(R,S)$
coordinates are depicted in
\cref{fig:coor}(e,f). Here the front connects $S_-$ with
$S_+$ while $R=R_*$ is constant.

\begin{figure}
\begin{center}
\includegraphics[width=\textwidth]{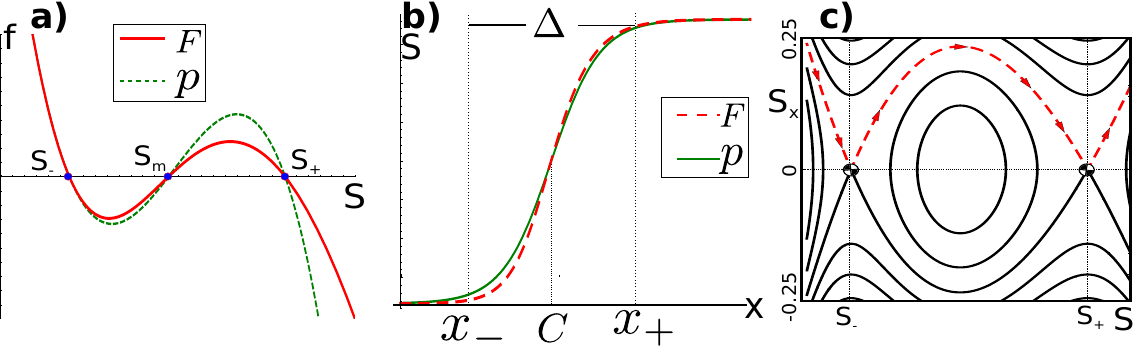}
\end{center}
\caption{Energetic description of wave pinning. (a) The function $F$ 
in terms of the new variable $S$, and its cubic approximation $p$. 
(b) Graph of a front (heteroclinic) solutions; 
numerical solution of \cref{wpnv:S} (dashed line) and analytical solution 
\cref{frentecub} (solid line). (c) Phase space obtained from 
level sets of the energy $E$, with the dashed red line corresponding 
to the heteroclinic solution connecting $S_-$ with $S_+$. 
Parameter values are: $R_*=1.849,\delta=0.06,\gamma=15,\eta=15, S_-=-1.8273,S_m=-1.77756, S_+= -1.71962$}
   \label{fig:big_pic}
\end{figure}

The right-hand side of equation \cref{wpnv:S} can be seen as the
derivative of a potential. Thus, the system conserves the
energy given by
\begin{equation}
\label{energia}
E=\frac{1}{2} \left(\frac{d S}{dx}\right)^2+V(S,R_*).
\end{equation}
Now, energy arguments can be used to construct solutions, because
values of $S$ that corresponds to minima of the potential $V(S,R_*)$, represent equilibria that
are spatially stable. 
When both minima of the potential have the same value, the system is
said to be at the Maxwell point \cite{Meronbook}.
Here, a heteroclinic cycle exists between the two
equilibria. This solution represents a front solution (and its corresponding
`back'). We shall call the energy value at which such fronts and
backs exist as $E=E_f$. 
Given $E_f$, it is possible to solve the differential equation \cref{energia} by separation of variables and obtain an implicit 
expression for the front solution
\begin{equation}
\label{eq:frente}
\int dx=x+C =\int \frac{dS}{\sqrt{2(E_f-V)}}.
\end{equation}
The family of front solutions are parametrised by $C$, which represents
the position of the core.  
Considering the kinetics given by \cref{dimlecz}, 
for a certain values of parameters, the system \cref{wpnv:S}
has a Maxwell point and it posses three zeros $S_-,S_m,S_+$ (cf. \cref{fig:big_pic}(a)). 

Upon integration of the right-hand side of \cref{wpnv:S}, it is possible to
obtain a closed-form expression for the potential
\begin{align*}
V(S,R_0)&=2 \int F(S,R_0)dS+ V_0\\
&=\frac{3 \gamma  \delta  R_0^2+4 \gamma  \delta ^2 \left(\delta  \log
    \left(4 \delta ^2+(R_0+S)^2\right)-2 R_0 \arccot \left(\frac{2
        \delta}{R_0+S}\right)\right)}{2\delta}\\
&+\frac{2 R_0 S (\gamma  \delta +\delta -\eta )-S^2 (\gamma  \delta +\delta +\eta )}{2 \delta }+V_0.
\end{align*}
 The set of
equienergetic curves (including the heteroclinic orbit) in the
$(S,S_x)-$phase plane form the conservative phase portrait depicted in
\cref{fig:big_pic}(c). The numerical front
solution is highlighted with a red dashed curve in the space and
phase space in \cref{fig:big_pic}(b) and (c)
respectively. Even though this numerical solution satisfies expression
\cref{eq:frente}, obtaining a closed-form expression is cumbersome. 
Instead
we can consider $p$, a cubic polynomial approximation of $F$ given by
$$F\left(w=\frac{R_*+S}{2\delta}\right)=F(w)=\frac{p(w)}{1+w^2},$$
where:
$$p(w)=\frac{R_*}{w_+ w_-w_I}(w-w_i)(w-w_-)(w-w_+).$$
In  \cref{fig:big_pic}(a), a comparison between $F$ and $p$ is
depicted (red continuous and green dashed lines respectively). The
cubic approximation $p$ presents qualitatively the same behaviour as
$F$. Considering $p$ it is possible to solve \cref{eq:frente} analytically and obtain a closed-form expression for the front
\begin{subequations}
\label{sol:cons}
\begin{align}
\label{frentecub}
S(x)&=S_-+\left(\frac{S_+-S_-}{2}\right)\left(1+ \tanh [\xi(x-C)] \right),\\
\label{frontscaling}
 \xi &=\left(\frac{S_+-S_-}{4}\right) \sqrt{\frac{2 R_*}{(R_*+S_-)(R_*+S_m)(R_*+S_+)}}.
\end{align}
\end{subequations}

The expression \cref{frentecub} captures the essential features of the
front-like solution. In particular, \cref{frentecub} provides an
analytical approximation of the ``width'' of the front ($\Delta$ in
\cref{fig:big_pic}(b)).  The values $x_{\pm}$ correspond approximately
to the extreme
values of the second spatial derivatives of \cref{frentecub}. The
width of the front an its relation with the core position are given by
\begin{equation}
\label{anchofrente}
x_{\pm}=C\pm \frac{\Delta}{2}, \quad \Delta=\frac{2}{\xi} \mbox{arcsinh}  \left(\frac{1}{\sqrt{2}}\right).
\end{equation}

\section{Localised states of the non-conservative system}
\label{sec:resultados}

In this section we now consider an investigation of the dynamics 
of \cref{dimle} in the case $\eps=1$. As a first remark, in contrast with the case where the mass is conserved, this new system posses a unique homogeneous equilibrium  given by
\begin{equation}
(u_0,v_0)=\left(\frac{\alpha}{\theta},\frac{\alpha (\varepsilon \theta +\eta)(\theta^2+\alpha^2)}{\theta[\theta^2+\alpha^2(1+\gamma)]}\right)=\left(\frac{\alpha}{\theta}, \beta_0+\varepsilon \beta_1\right).
\label{homoeq}
\end{equation}

\subsection{Linear and weakly non-linear stability analysis}
\label{sec:calcula}

Performing a linear stability analysis around the homogeneous equilibrium, we
can find the conditions for a pattern formation instability (also known
as a Turing bifurcation or, in spatial dynamics as a Hamiltonian-Hopf
bifurcation). Such bifurcations are of codimension-one in the parameter space,
see e.g.~\cite{crossrev,Murray2,Victor2}. 
To look for such bifurcations, we substitute the following ansatz into \cref{dimle}
\begin{equation}
\label{eq:linper}
(u,v)=(u_0,v_0)+(\bar{u},\bar{v}) e^{i k x-\sigma t} \quad ||(\bar{u},\bar{v})||\ll 1,
\end{equation}
neglect the non-linear terms for $(\bar{u},\bar{v})$ 
and impose that the maximum of the real part of $\sigma(k)$ be zero. 
This then allows us to solve for the critical parameter value and  
predict the 
critical wavelength of the bifurcation. 
Proceeding in the usual way, we find 
the expressions for the critical point and the wavevector to be:
\begin{subequations}
\label{criticas}
\begin{align}
\label{criticaa}
\varepsilon \theta \partial_v F -\frac{ (\partial_u F -\delta \partial_v F -\varepsilon \theta)^2 }{4\delta} &=0,\\
\label{critik}
k_c ^2=\frac{\partial_u F-\delta \partial_v F-\varepsilon \theta}{2\delta} &>0,
\end{align}
\end{subequations}
where $\partial_{\xi} F= \left.\frac{\partial F}{\partial \xi}\right|_{(u_0,v_0)}$ when $\xi=u,v$. 

We can understand the behaviour of the system in the vicinity of the spatial instability by introducing the change of variables 
\begin{equation*}
(u,v)=(u_0,v_0)+(U,V),
\end{equation*}
into \cref{dimle} to obtain 
\begin{equation}
\label{eq:equni}
\partial_t \left(\begin{array}{c} U\\ V \end{array} \right)=
[\mathbb{J}+\mathbb{D}\partial_{xx}]_c\left(\begin{array}{c} U\\ V \end{array} \right)+\left(\begin{array}{c}
1\\ -1\end{array}\right) \mathbb{NL} (U,V),
\end{equation}
where $[\mathbb{J}+\mathbb{D}\partial_{xx}]_c$ is the linear operator evaluated at the critical point
(evaluated at \cref{homoeq,criticas}), composed of the jacobian matrix
$\mathbb{J}$ and  the diagonal matrix accounting for the second
spatial derivative $\mathbb{D}$. The scalar quantity $\mathbb{NL}$ corresponds to the nonlinear terms in the expansion.

If we assume the pattern to be a time-independent linear solution of
\cref{eq:equni}, we obtain
\begin{equation}
\label{eq:linsol}
\left(\begin{array}{c}  U\\ V\end{array}\right)_{l}=(A e^{ik_c x}+\bar{A} e^{-ik_c x})\left(\begin{array}{c} \partial_v F\\ -\partial_u F +\varepsilon \theta +k_c^2 \delta\end{array} \right),
\end{equation}
where $A$ stands for the amplitude of the pattern. 
We are interested in finding an amplitude equation for 
$A$ in a neighbourhood of the instability. This amplitude equation can also
be thought of as the normal form of the spatial dynamical system
(where $x$ is thought of as a time-like variable), see
\cite{ioosbook}. 
The calculation of the coefficients of the normal form can be carried
out using the procedure outlined in \cite{normalforms}. 
The calculation
is lengthy but standard, and we relegate the details to the \cref{ap:nf}, giving
only the bare essentials here.  

The change of
variables and the amplitude equation obey the
ansatz:
\begin{subequations}
\label{eq:fnans}
\begin{align}
\label{eq:fnanscv}
\left( \begin{array}{c} U\\ V\end{array}\right) &=W^{[1]}+W^{[2]}+\ldots,\\
\label{eq:fnansa}
\partial_t A&=\partial_t A^{[1]}+\partial_t A^{[2]}+\ldots,
\end{align}
\end{subequations}
where the superscript accounts for the order in $A$. 
We can solve this equation at each order; for example, at first order we have:
$$
\partial_t W^{[1]}=\partial_{A}W^{[1]}\partial_t A^{[1]}
=[\mathbb{J}+\mathbb{D}\partial_{xx}] U^{[1]}.
$$
The choice $\partial_t A^{[1]}=0$ reduces this equation to a linear one 
and therefore we have 
$$
W^{[1]}=\left(\begin{array}{c}  U\\ V\end{array}\right)_{l} .
$$
At second order we can choose $\partial_t A^{[2]}=0$ as well. 
Actually this is possible for \emph{every even order}. 
On the other hand, when dealing with odd powers, 
there will be \emph{resonant terms} 
(terms which are proportional to \cref{eq:linsol}, 
the vectors in the kernel of the linear operator \cref{eq:equni}) 
on the right-hand-side of \cref{eq:equni}. 
In order to ensure that the problem at each odd order is solvable, 
we need to impose a \emph{solvability condition} using the 
Fredlholm Alternative theorem. 
In summary, we find an amplitude equation of the form
\begin{equation}
\partial_t A=\epsilon C_1  A +C_3 A |A|^2+C_5 A |A|^4+
\mathcal{O}(|A|^6 A). 
\label{amplieq}
\end{equation}
Here $\epsilon$ is an unfolding parameter that accounts for the parameter variation around the critical point. The constants $C_i$ are obtained 
from the solvability conditions and 
are functions of the parameters evaluated at the bifurcation point. 
More details are given in the \cref{ap:nf}.

 In our study, the main purpose for computing the amplitude equation is to find where the
bifurcation changes from being subcritical to supercritical. As argued in \cite{Victor2}, sub-criticality of the Turing bifurcation
is a necessary ingredient for the birth of localised structures in
reaction diffusion systems.  
 This such a transition point represents the nascence of a bistability region, where the homogeneous
state coexists with the patterned one. 
That transition happens whenever $C_3$ changes sign in equation \cref{amplieq} . 

Formally speaking, for the amplitude equation, we also need to check
that $C_5<0$ in order to ensure the existence of a higher order coefficient which stabilises the
amplitude equation. We have found evidence for this numerically.

Considering $C_3=0$ and \cref{criticaa} as implicit functions of two 
of the system parameters, the nascence of bistability 
will thus occur at the codimension-two intersection point of both curves.

\subsection{Numerical simulation results}
\label{sec:numinv}

The system \cref{dimle} with $\eps=1$ has five parameters, and it is
infeasible to explore the effect of varying every possible
combination.  So, following \cite{Mare,Morior,revjilkine}, we fix all
of them except two and consider the effect of variation only of the
non-linearity in the system $\gamma$ and the diffusion ratio $\delta$. The fixed values of the other parameters will
be taken to be
\begin{equation}
\varepsilon=1, \quad \eta=5.2, \quad \theta=5.5, \quad \alpha=1.5, \quad 
L=100,
\label{parvalues}
\end{equation}
unless otherwise stated. 

\begin{figure}
\begin{center}
\includegraphics[width=\textwidth]{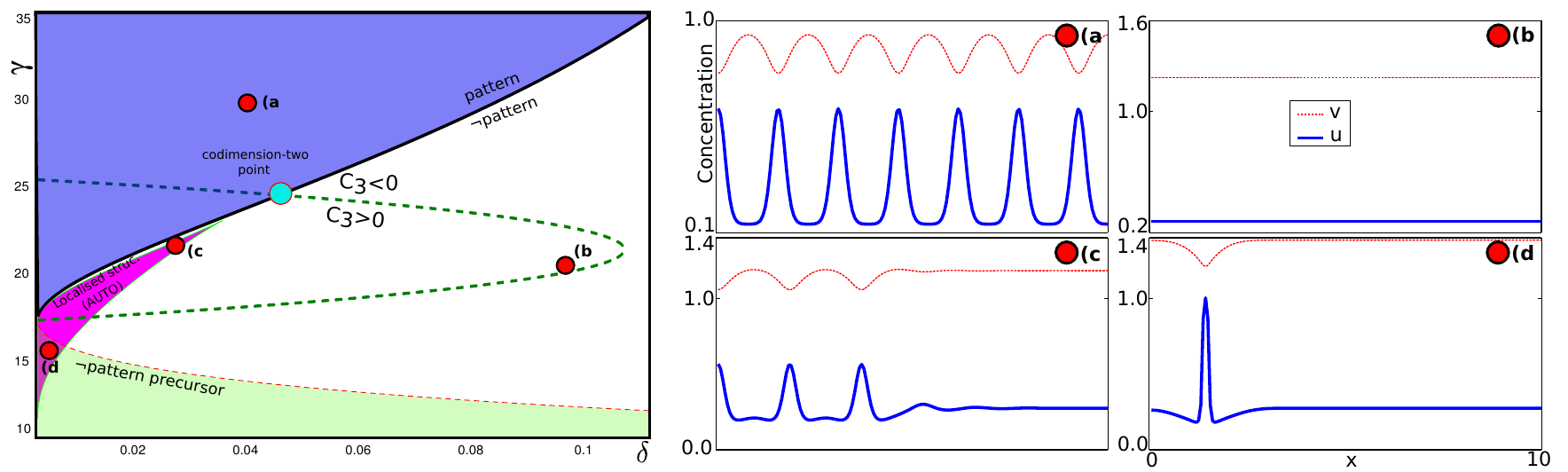}
\end{center}
\caption{Numerical two-parameter bifurcation diagram for the 
model \cref{dimle} with fixed parameters \cref{parvalues}
The pink shaded region is where localised solutions exist.  
The labelled points correspond to the qualitatively
different stable equilibrium solutions illustrated in the corresponding
sub-figures. The $(\delta,\gamma)$-values depicted are: 
(a) periodic (Turing) pattern at $(0.0406,30)$, 
(b) homogeneous equilibrium at $(0.10272,20)$,  (c) localised pattern at 
$(0.02688,21.2411)$ and (d) a lone spike solution at $(0.002,14.8)$.} 
\label{fig:figura2}
\end{figure}

The left-hand panel of \cref{fig:figura2} shows the basic bifurcation
curves in the $(\delta,\gamma)$-parameter plane, in a region where all
the qualitatively different behaviours can be observed. The black
continuous line corresponds to the spatial instability curve (equation
\cref{criticaa} when $k_c^2$ is positive in \cref{critik}). This
curve splits the parameter space into a region where just patterns are
observed (above the line) and the rest. The dashed green line
indicates where the cubic coefficient in the amplitude equation
\cref{amplieq} vanishes. This curve is only relevant at its
intersection with the spatial instability curve, at the light blue
point. For values of $\delta$ smaller (greater) than the light blue
point, a \emph{sub-critical} (\emph{super-critical}) bifurcation for
the amplitude of the patterns takes place. The red dashed line in the lower
part of the figure indicates where the dispersion relation
\cite{crossrev} of the homogeneous states changes from having a
maximum at zero (beneath this line) to have a non-zero maximum
(above), the points in this curve correspond to a bifurcation which
was termed a \textsl{Belyakov-Devaney} (BD) point in \cite{apd}.

All the curves and points listed so far were obtained through analytic
calculations. Moreover, the pink region corresponds to the region
where localised structures were found using numerical continuation
(the details will be explained below). Finally, the red dots are
representative points in the parameter plane, where qualitatively
different solutions are observed.

In order to explore the dynamics of model \cref{dimle}, we proceed to
simulate the full PDE system using finite differences in space (with
$dx= 0.01$) and fourth-order Runge-Kutta scheme in time (with fixed
timesteps $dt=10^{-3}$). Appealing to the Neumann boundary conditions, we also simulate on a
half-interval $[0,L]$, with results on the full interval
being obtainable by reflection in $x=0$. Sweeping parameters and
trying different initial conditions, it is possible to corroborate the
predictions made by the linear stability analysis (\cref{sec:calcula})
and also find localised structures. The insets of \cref{fig:figura2}
on the right, depict the representative solutions.

The solution in \cref{fig:figura2}(a) is a Turing pattern whose
region of existence can be predicted using the critical condition
\cref{criticas}.  These patterns can be characterised in terms of
their wavelength and amplitude which can be obtained using expressions
\cref{critik} and \cref{amplieq}.  Note how the oscillations of the
  active and inactive states are in anti-phase.  These predictions are
  only valid in the vicinity of the instability.  The homogeneous
  state observed in panel (b) is given by \cref{homoeq}.

Further, the solution in \cref{fig:figura2}(c) corresponds to a
localised patterned state, which when reflected onto the full domain
$[-L,L]$ has five localised peaks of the active
state (which correspond to troughs of the inactive state). In this
parameter region we can find such localised patterns with an arbitrary
number of peaks. Such solutions arise because there is a heteroclinic
connection between the homogeneous state and the Turing patterns (as
we shall explain shortly using the theory of homoclinic snaking).
Note that there is a  wavelength of the Turing patterns
which corresponds to a well-defined distance between each of the
peaks.  These localised states coexist with the stable
homogeneous, flat state.  Sufficiently small initial conditions tend to
converge to the flat solution whereas arbitrary sufficiently large
initial conditions tend to converge to localised patterns with the
numbers of peaks depending on the precise features of the
initial data. On very long domains we can also find multiple localised
patch patterns that are separated by long intervals of (almost)
homogeneous solution.

The solution depicted in \cref{fig:figura2}(d) is also a localised
solution, but is rather different. Here we see a single isolated peak
(trough).  This pattern has no underlying wavelength and the decay of
the tails of the peak are monotonic rather than oscillatory. Here
initial conditions are found to converge to either the stable
homogeneous state or to just these single isolated peaks. If
multi-peaked initial data is used, then over a short timescale,
several well-separated peaks can be be formed. These peaks are then
found to separate at a speed that decreases (exponentially) as the
separation increases. Eventually, either the subsidiary peak is
annihilated or it disappears to the edge of the domain, or, due to
numerical noise a bound state can be formed consisting of two or more
peaks separated by a large interval of homogeneous state. In fact,
taking the mirror image of the solution in \cref{fig:figura2}(d)
we would have just such a ``numerical'' bound state.

Note that these two distinct kinds of localised solutions, either
localised patterns with an underlying wavelength, or single isolated
peaks are present in a broad spectrum of partial differential
equations, derived in several different contexts; see \cref{sec:concl}
for a discussion.  In our case, a definite point of transition between
the two behaviours can be identified (the red line in the lower region
of \cref{fig:figura2}), which will be we explain in the next
subsection.

\subsection{Numerical continuation results}

From the direct numerical simulations, it is possible to
conclude that despite the system \cref{dimlex} not being variational, it
does not present permanent dynamic behaviours such as limit cycles or
chaos (at least not in the parameter regime 
under consideration). Therefore, 
we can focus our attention on time-independent
solutions. This assumption reduces the partial differential equation to
a reversible 4-dimensional systems of ordinary 
differential equations (ODEs), which can be studied using 
{\em spatial dynamics}, see e.g.~\cite{alansnake}. 
Specifically, we have
\begin{equation}
\partial_x \left(\begin{array}{c}a\\ \partial_x a\\  b\\ \partial_x b\end{array}\right)=\left(\begin{array}{c} \partial_x u\\
\frac{\varepsilon u -F(u,v)}{\delta}\\
\partial_x v\\
F(u,v)-\varepsilon \alpha \end{array}\right).
\label{spadyn}
\end{equation}

Among the many advantages of studying the ODEs \cref{spadyn} instead of 
\cref{dimle}, is that the we can easily perform numerical continuation
to look for periodic and localised solutions, for instance, 
using the software \verb#AUTO# \cite{auto}. Numerical continuation allows 
us to unveil the region of existence, bifurcation and transition
of the different solutions present in the system. Nonetheless, 
there are a few drawbacks. The continuation results
do not give information on temporal stability, for that we need to
study the full PDE system via simulation or spectral computation. 

Taking the \emph{localised pattern solution} from \cref{fig:figura2}(c)
as a starting point for the continuation in $\gamma$, one obtains a
sequence of solutions for the different values of $\gamma$. In order
to visualise the continuation, it is useful to represent some
unidimensional quantity as a function of the parameter of
continuation. One possibility is what we term $L_2-Norm$ given by:
$$L_2=\sqrt{\frac{1}{L}\int_0^{L} a(x)+\partial_x a(x)+
  b(x)+\partial_x b(x)
dx}.$$

\begin{figure}
\begin{center}
\includegraphics[width=7.5cm]{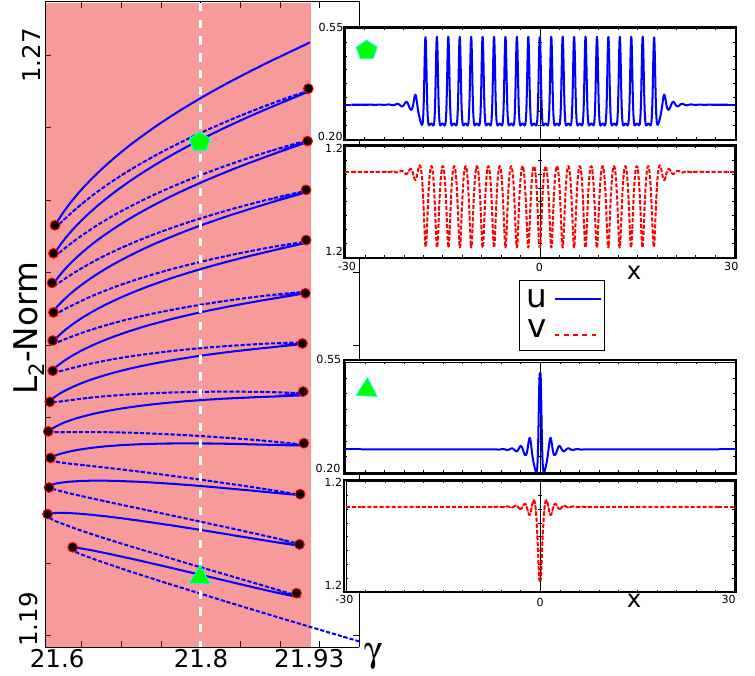}
\end{center} 
\caption{Numerical continuation in $\gamma$ of the localised pattern solution 
of \cref{dimle} for $\delta=0.02688$ and other parameters as 
in \cref{parvalues} $\eta=5.2$, $\varepsilon=1$,  
$\theta=5.5$ and $\alpha=1.5$. 
(Left) $L_2-$norm versus $\delta$; 
dashed (continuous) lines represent unstable (stable) solutions.
Each stable--unstable pair of solutions is connected through a saddle-node 
bifurcation marked by a black dot. (Right) Two distinct solutions for 
$\gamma=21.8$ (marked by a triangle and a pentagon on the left-hand panel)}
\label{fig:snake}
\end{figure}

In the \cref{fig:snake}, curve shown following by the path is
presented when the $L_2-Norm$ is considered.
The most remarkable feature of this graph is the switching back and
forth in a behaviour known as the \emph{homoclinic snaking}
\cite{alansnake,burke01,snakebeck,Victor3}.  Within a certain region
of the parameters, a family of solutions can be observed. Each one of
these solutions exist in a line connecting two turning points, where a
saddle-node bifurcation occurs. Hence, the two branches connecting any
of these points have different stability. In order to determine the
stability of each solution, we replace the different solutions in the
full PDE \cref{dimle} system and evaluate their persistence under
perturbations.  Here stable (unstable) solutions are represented by continuous (dashed)
lines.  \cref{fig:snake} also shows two different stable
solutions that exist for the same value of $\gamma$. Note that these
solutions are left-right symmetric. According to the theory of snaking
in non-variational systems (see \cite{Knoblochreview}), there will be
a bifurcation of a branch of non-symmetric states close to each
saddle-node, but these will represent travelling rather than
stationary states. Any stationary localised structure must be
symmetric.  The series of saddle-node bifurcations occur approximately
at two values of $\gamma$, defining an interval known as the
\emph{snaking region} or \emph{pinning region} where here pinning
refers not to the front solutions of the wave-pinning model but to the
pinning of the heteroclinic connection between the homogeneous state
and Turing pattern. However, owing to the snaking region extending
into the pulse region, we shall refer to the region as being the 
\emph{localisation region}.
The pink rectangle in \cref{fig:snake} represents
the extent of the localisation region. 
Note that qualitatively similar results are
obtained if we continue in $\delta$ rather than $\gamma$ from the same
initial localised state.

Using two-parameter continuation to trace the saddle-node 
points as two parameters vary,
it is possible to delineate the snaking region 
in the $(\delta,\gamma)$-space. Tracking one fold at each edge of the
snaking interval, we obtain two curves in the
$(\delta,\gamma)-$space. The area delimited between these lines
corresponds to the localisation region, which is also indicated in pink
in the two-parameter plot in the left-hand panel of 
\cref{fig:figura2},

In theory, we should expect the localisation region to extend all
the way to the subcritical Turing bifurcation point. But note from 
\cref{fig:snake} that that the portion of the localised solution from
the Turing bifurcation to the first fold is unstable. As we approach
the codimension-two super-to-subcritical transition point, then, in accordance
with theory \cite{Kozyreff,Knoblochreview} the localisation region becomes an 
exponentially thin wedge which proceeds algebraically from the
codimension-two point.  

In practice, because of numerical precision and the finite domain effects
(see \cite{Dawes})
the numerical routine used to follow the saddle-nodes into the
exponentially thin region breaks down. However the location of the (Maxwell) line
in the centre of the localisation region can be easily located
numerically. 

\begin{figure}
\begin{center}
\includegraphics[width=.8\textwidth]{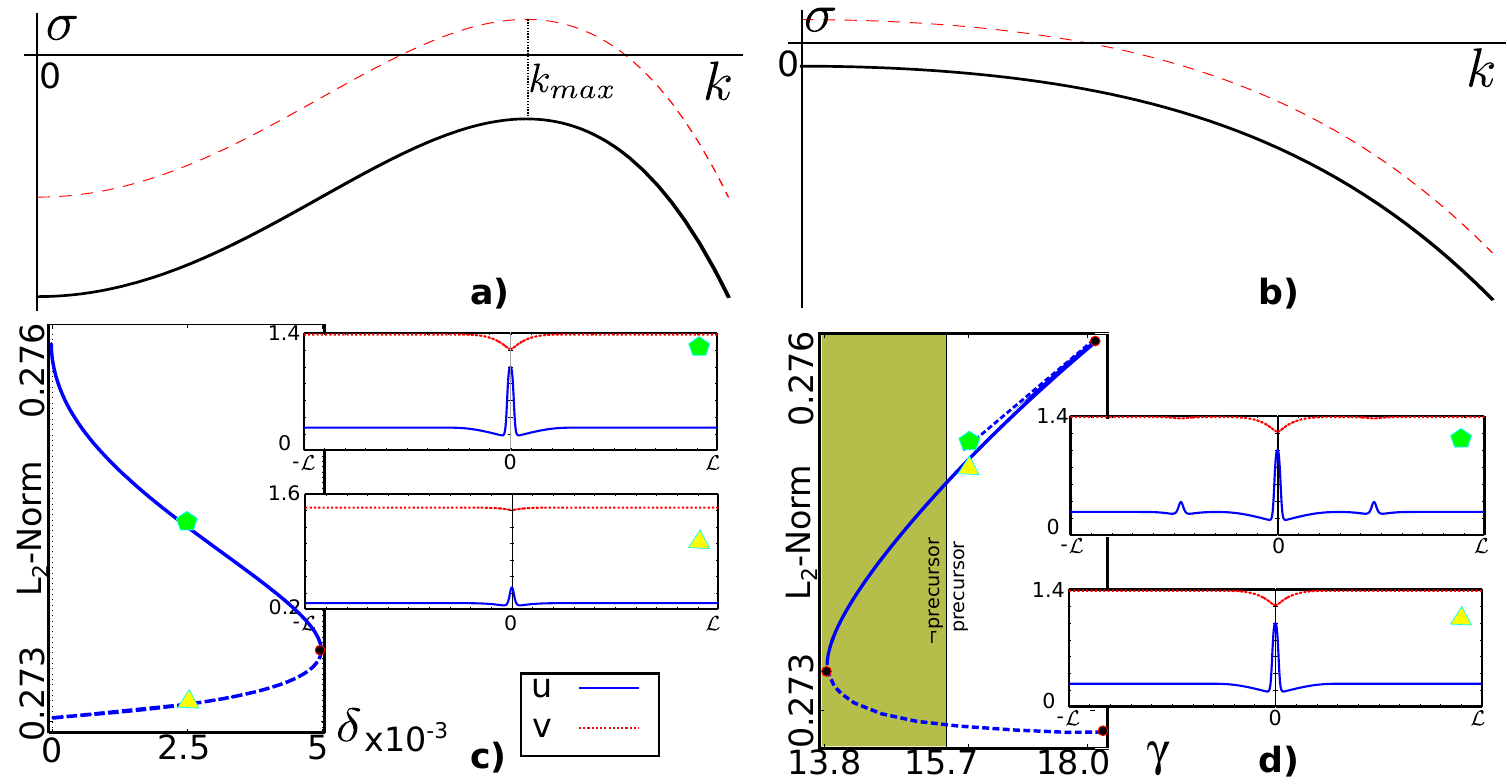}
\end{center}
\caption{(Top) Sketch of the dispersion relation $\sigma(k)$ for (a) 
type I and (b) type III. The dashed red (black continuous) curve 
corresponds to the unstable (stable) case. In \textbf{a)} $k_{max}\neq
0$ and therefore there is a precursor. (Bottom) One-parameter continuation results 
for the spike solution. When $\delta$ is varied (c), 
one saddle node bifurcation connects the large (stable) and 
small (unstable) solutions. When $\gamma$ is varied (d), 
two saddle-nodes are observed.  In the region where there is 
no pattern precursor (dark shaded panel), the saddle-node 
plays the same r\^ole as in (c). The second saddle-node connects to a
branch of solutions with addition subsidiary peaks different number of maxima, 
illustrated in the insets on the right.}
\label{fig:spikont}
\end{figure}

The two-parameter plot reveals the richness presented by the family of
localised solutions. In particular, both the spike and
localised patterns solutions are both contained in the localisation region (see the left panel of 
\cref{fig:figura2}). In order to understand the mechanism of transition between
these two kinds of state, we perform a one-parameter numerical continuation 
of the spike solution. The results are 
presented in \cref{fig:spikont}.
In contrast with the localised patterns, for the spike solution
continuation in $\delta$ and $\gamma$ leads to qualitatively
different results. As can be seen in
\cref{fig:figura2}, the continuation of the spike 
solution in $\gamma$ crosses the line which represents the BD transition,
whereas continuation in $\delta$ does not cross this line. 

According to linear stability analysis (cf.~\cref{sec:calcula}), 
a homogeneous equilibrium will be linearly stable (linearly unstable) 
if the maximum of the real part of the dispersion relation $\sigma(k)$ 
is negative (positive).  At zero, we are in the critical situation 
(e.g.~\cref{criticas} in our case). 
The dispersion relation can exhibit two qualitatively different behaviours, 
namely type I and III in the notation of \cite{crossrev}, 
corresponding to the maximum for $k$ being at zero and non-zero respectively
(see \cref{fig:spikont} (a) and (b)). In a stable regime (max$(\sigma(k))<0$), any
perturbations to the system will be decompose into Fourier modes with 
different wave numbers whose maximum is $k$, 
the \emph{slowest} decaying mode. 
As a consequence, if we are in the case I, a non-zero wavelength will be 
observed during the transient dynamics that approaches the steady state 
solution. In this case it is said that there is the existence of a 
\emph{pattern precursor}. On the other hand, a instability type III 
will not exhibit a precursor. Hence, in the left panel of \cref{fig:figura2}, 
the BD transition point corresponds to the point of 
transition between dispersion curves of 
type I and III. 
This transition curve can be determined analytically; 
in the case of \cref{dimle}, it corresponds to \cref{criticaa} 
when \cref{critik} is negative. 

\cref{fig:spikont}(c) shows the continuation in $\delta$.  
Here the saddle-node bifurcation corresponds to the right-hand limit 
of the localisation region, and connects the stable single spike
solution to a lower-amplitude unstable single spike, as illustrated
in the insets.

In contrast, \cref{fig:spikont}(d) shows continuation in $\gamma$. 
Here the solution crosses the BD line where the nature of  
$\sigma(k)$ changes from type I to type III. (cf.~\cref{fig:spikont}(a)
and (b)). In the absence of a pattern precursor (in the dark green shaded
region) the solution is analogous to that with continuation in $\delta$. 
Outside of this region, the spike solution develops oscillatory tails, 
and we are in the parameter region where we would expect to see
homoclinic snaking. As part of the snaking scenario, the 
switching back through the right-hand saddle-node corresponds to 
where the solution acquires a new (symmetric pair of) localised peaks.
As this new three-peak solution is traced back towards the BD point, the 
continuation fails to converge before reaching there. It is clear what
 happens is that the wavelength of the precursor tends to infinity
as the BD point is approached and so the separation between the the three
peaks in the multi-peaked state also becomes infinitely large. Hence
the branch of multi-peaked solutions disappears in a non-local bifurcation
at the BD point. See \cref{sec:concl} for further discussion on
this non-local bifurcation. 
 
Thus we can see that the BD line splits the localisation region
into two parts, the snaking region where localised patterns exist,
and the region where there are only isolated spike solutions.
Moreover, this observation is in complete accordance with the theory
in \cite{Claudiohole} (and references therein), in which the existence
of a spatial wavelength in the system is claimed to be necessary
ingredient for the existence of localised patterns.

\section{The mass-conservation limit $\varepsilon \to 0$}
\label{sec:consv}

So far we have been investigating how the wave-pinning model changes
when the mass is not conserved. Since in a realistic scenario the
total mass is \emph{roughly} constant, we are interested in the case
of small $\varepsilon$. From a mathematical point of view, it is
also intriguing to ask how the structure of localised and extended
patterns collapses to the wave-pinned solutions as we pass 
to the mass-conservation
limit $\varepsilon \to 0$. In particular, wave-pinning
naturally leads to front solutions (heteroclinic orbits), a
solution where the spatial symmetry is broken, which can
be argued to be a necessary ingredient for cell polarisation. 
 Indeed, when we run simulations for $\eps=0$, inhomogeneous 
initial conditions on long domains quickly form a state with a number of fronts and 
backs which tend drift and coarsen into
a single wave-pinned front. For $\eps=1$, on the other hand, in the 
localisation region inhomogeneous initial
conditions on long domain quickly form states with a number of spikes (or localised
patterns) which slowly drift into either isolated spikes (or localised patterns).  
We now seek to investigate how these two very different kinds of long term dynamics
can be connected as we vary $\eps$ from 1 and 0. 
 
\begin{figure}
\begin{center}
\includegraphics[width=11cm]{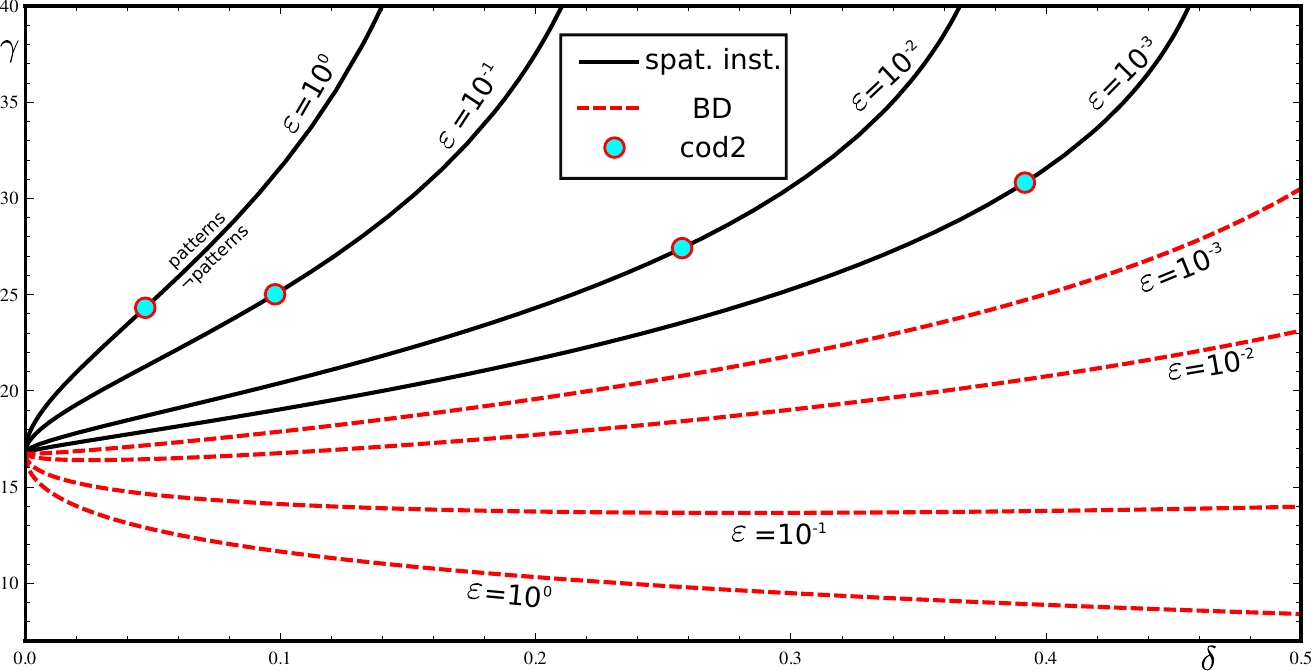}
\end{center}
\caption{Local bifurcation curves in the ($\delta,\gamma$)-plane for
different values of $\varepsilon$.  
The continuous black lines correspond to the Turing (spatial
instability) bifurcation and the dashed red line
to the BD transition point. In light blue the codimension-two point where the nascence
of localised structures takes place.}
\label{fig:spikeregion}
\end{figure}
As a first step in understanding the mass-conservation limit, in
\cref{fig:spikeregion} we have plotted the two critical curves where a
Turing instability and a BD transition take place (given by the
analytic expressions \cref{criticaa} when \cref{critik} is positive and negative
respectively), for a variety of values of $\varepsilon$. Approaching the limit $\eps \to 0$, we
find that these two curves approach each other. In fact it is possible
to show that these curves become identical as $\eps \to 0$. 
Hence, if any localisation region survives into the limit $\eps\to 0$, the 
 region of localised patterns will vanish, so that we are only left with
localised spikes in the limit.

\subsection{Numerical results}
\label{sc:mcnr}
\begin{figure}
\begin{center}
\includegraphics[width=\textwidth]{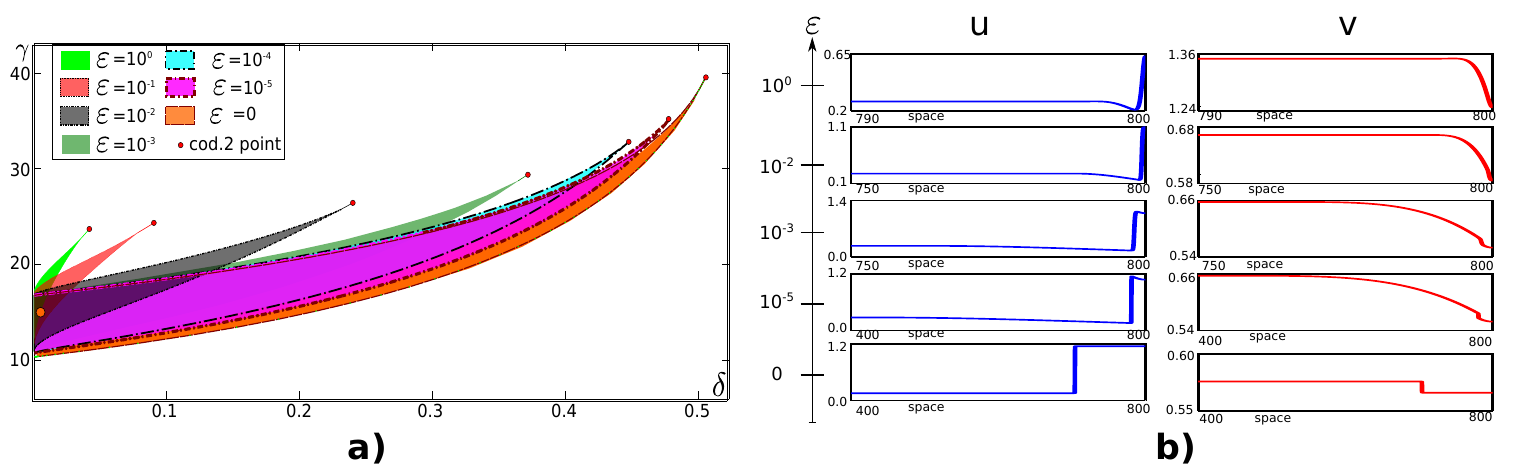}
\end{center}
\caption{(Left) comparison of the different \textsl{localisation region}
    as $\varepsilon$ is varies. The red points correspond
  to the codimension-two super- to sub-critical Turing bifurcation transition point. 
Note the point identified in orange (at $(\delta,\gamma)=(0.01,16.8)$) which 
lies in the localisation region for all $\eps$ values depicted. 
(Right) Solution of the pulse solution at this point for
different $\eps$ values, plot on the half-domain $[-L,0]$. The
parameter values are the same as in \cref{parvalues}, except for $L=800$.
} 
\label{fig:snakep}
\end{figure}

The natural next step is to compute, using the same procedure 
as in \cref{sec:numinv}, the localisation region for different values of $\varepsilon$. 
The results are presented in \cref{fig:snakep}.
At first glance, notice how the localisation region grows as
$\varepsilon$ tends to zero. Hence, according to
\cref{fig:spikeregion}, this region becomes increasingly composed of
spikes rather than localised patterns. 
 Interestingly, the intersection of this region with $\delta=0$ seems to be
independent of $\varepsilon$.

In order to compare the equivalent of the localisation region 
at $\eps=0$, we have delineated precisely the 
region in the two-parameter space where the wave-pinned fronts exist.
Since the spatial dynamics of model \cref{dimle} is singular at $\varepsilon=0$, 
we can consider the equivalent 2-dimensional problem obtained 
using the $(R,S)$ variables
(see equation \cref{wpnv} in \cref{sec:energy}). 
Imposing the continuity between the different localisation regions,
the extra parameter $R_*$ in \cref{wpnv} is chosen such that 
$R_*=\lim_{\varepsilon \to 0}\delta u_0+v_0 $, where $u_0,v_0$ are given by \cref{homoeq}.
The codimension-two point at the tip of the region, which is calculated
from the limit $\varepsilon \to 0$ in the analytical expression for
$C_3$ in \cref{amplieq}, gives a perfect prediction
of the tip of the front region region obtained by two-parameter continuation.
Moreover, although the localisation region grows significantly, it appears
to vary continuously as $\eps$ is reduced to zero, with the upper and lower
bounds of the front region continuing naturally into the upper and lower saddle-node
bifurcations of the spike solutions for non-zero $\varepsilon$.

A next question is how localised spike solutions change their 
shape into being a \emph{front}.  \cref{fig:snakep}(b) illustrates how graphs 
of the solution are transformed as $\eps$ is reduced for fixed
values of $\delta$ and $\gamma$. We start from the spike solution, in the middle of the 
domain, which we compute on the half-domain. 
As $\varepsilon$ gets smaller, both components $u$ and $v$ start to
develop a significant shelf so that the solution on the full domain 
resembles a front and back pair. Looking at just the $u$-component, we might imagine
that the pulse converges uniformly to this front and back pair. Note also
that, as expected, the exponential decay in the tail becomes progressively weaker.

However, there appears to be a subtlety. Looking at the penultimate solution,
depicted for $\eps=10^{-5}$ , we note that on the domain size depicted,
although the decay rate of the tail gets weaker, its amplitude if anything 
(especially in the $v$-component) appears to grow. Also, the core of the pulse
is not flat but appears to have a dimples in both $u$ and $v$ components.
When comparing with the corresponding front solution at $\eps =0$, we see that
the chosen left-hand limit of the solution is not the same as that of the front.
Instead the non-vanishing weakly decaying tail seems to play the r\^ole of
adjustment of the unique equilibrium values $(u_0,v_0)$ for the pulse 
to the asymptotic values of the selected front of the wave-pinning
model.  

We therefore turn to asymptotic analysis to explore this curious phenomenon
in the singular limit $\eps \to 0$. 

\subsection{ Asymptotic analysis}
\label{ssec:asym}

In the language of dynamical systems, the spike solution corresponds
to a homoclinic orbit of the spatial system and the front
solution to a heteroclinic one. Thus, considering the extended
domain $[-L,L]$, the problem can be reformulated as the transition
between a homoclinic solution whose maximum is at $x=0$ and a
heteroclinic loop. Since the solutions considered are even,
the attention can be restricted to the domain $x\in [-L,0]$.
In  \cref{sec:energy}, we showed in the conservative case 
how the spatial dynamics can be reduced from a 4-dimensional space phase 
into a 2-dimensional one with a free parameter $R_*$. 
Hence, the transition involves a reduction of the number of degrees of freedom of 
the system as well as the nascence of a unique homogeneous
equilibrium. 

In order to study this transition, in the spirit of \cref{sec:energy},
we rewrite the spatial system in terms of the $(R,S)$ variables. We
shall also remove the dependence of the homogeneous equilibrium (given by \cref{homoeq}) on $\varepsilon$. Specifically, let 
\begin{equation}
\label{eq:cve}
 u=\frac{R+S}{2\delta}, \quad v=\frac{R-S}{2}+ \varepsilon \beta_1.
\end{equation}
Defining
$$
\beta_1=\frac{\alpha(\theta^2+\alpha^2)}{\theta^2+\alpha^2(1+\gamma)},\quad \psi(u^2)=\frac{\gamma u^2}{1+u^2}+1,
$$
and substituting the new variables \cref{eq:cve} into the time-independent version of (\ref{dimle}), we obtain
\begin{subequations}
\label{sysas}
\begin{align}
\frac{d^2 R}{dx^2}&=\varepsilon \left[\frac{\theta}{2\delta} (R+S)-\alpha \right],\\
\frac{d^2 S}{dx^2}&=\varepsilon \left[\frac{\theta}{2\delta} (R+S)+\alpha-2\beta_1 \psi\left( \left(\frac{R+S}{2\delta}\right)^2 \right)\right]\\
\nonumber &-2F\left(\frac{R+S}{2\delta},\frac{R-S}{2} \right).\quad x\in ]-L,0].
\end{align}
\end{subequations}

\begin{figure}
\begin{center}
\includegraphics[width=\textwidth]{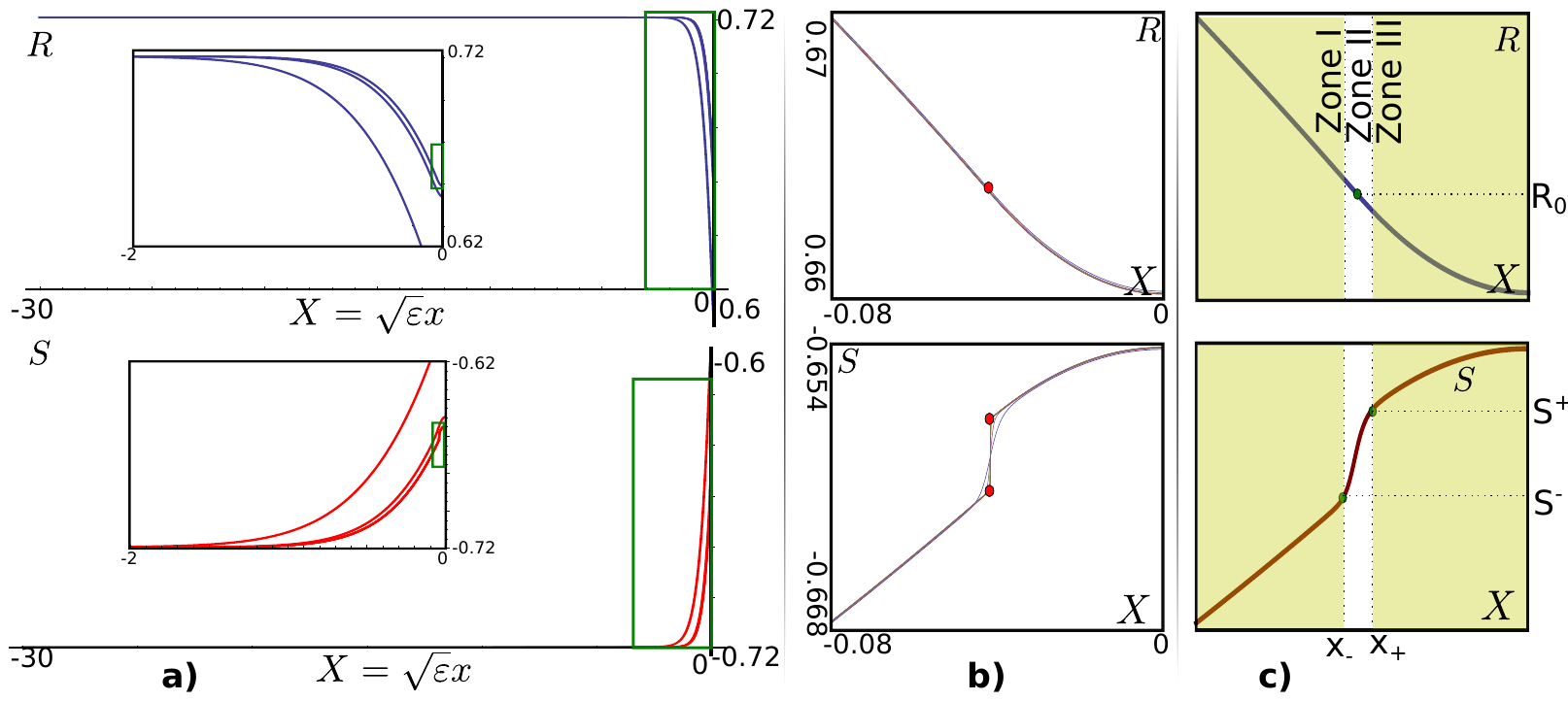}
\end{center}
\caption{Solutions of \cref{sysas} in the half domain $[-L,0]$ as
  $\varepsilon$ tends to $0$. The values of the parameters are:
  $\delta=2\times 10^{-3}, \gamma=14, \eta=5.2, \alpha=1.5,\theta=5.5,
  L=10^5$.  Superposition of the $R$ (blue) and $S$ (red) components
  of the homoclinic solution for $\varepsilon=\{10^{-n}\}_{n=0}^{9}$,
  the scaling $X=\sqrt{\varepsilon}x$ has being used in order to
  visualise all the solutions in the same plot. Zooms of the solution,
  highlighted in the green rectangle and shown in the insets. (b) Zoom
  around the front for both components. (c) Schematics of the
  behaviour for both components (see text for details). }
 \label{fig:solnv}
\end{figure}

We can study the transition on the new system \cref{sysas} by
performing one-parameter numerical continuation in $\varepsilon$. Our
starting point is the is the half-homoclinic solution at the orange point in 
\cref{fig:snakep}, transformed in the new variables \cref{eq:cve}. We will start our analysis by considering the case of 
the semi-infinite domain, with solutions that are  asymptote to the same homogeneous
value as $x \to -\infty$. To that end, we shall choose $L$ to be arbitrarily large
(we shall consider finite-domain effects in the next subsection). 
Consequently, the front is developed in a small region of the space,
making it difficult to visualise the dynamics for the different values of $\eps$.
It is then instructive to plot the solutions using the new scale

\begin{equation}
X=\sqrt{\varepsilon}x, \quad X\in ]-\mathcal{L},0], \quad \mathcal{L}=\sqrt{\eps}L.
\label{eq:scaling}
\end{equation}
In  \cref{fig:solnv}(a), ten solutions for representative values of 
$\varepsilon$ are superimposed  using the new scale. The solutions are hardly distinguishable as
$\varepsilon \to 0$. This is a consequence of the solutions converging
to a well-defined limit.

Zooming into the solutions, we observe the
appearance of a front-like behaviour in the $S$ component
(see the inset for $S$ in \cref{fig:solnv}(b)). This front solution can be studied as a \emph{interior layer} problem
using \emph{matched asymptotics} \cite{kevorkian,MHolmes}. In this
framework, two dynamical regimes can be distinguished. An \emph{inner}
zone where the solution
varies on the fast $\mathcal{O}(x)$ scale and an \emph{outer} zone,
where the solution varies in the slow $\mathcal{O}(X)$ (given by
\cref{eq:scaling}) scale. In the case of an interior layer, the space
can be divided into three zones, as indicated in 
\cref{fig:solnv}(c); two {\em outer} zones (I and III) divided by an
{\em inner} zone (II).\\

Assuming Neumann boundary conditions and imposing continuity of the
solution in the transitions between inner and outer zones, the
boundary conditions can be defined for each zone. In the \cref{tab:bc}, the zones
and boundary conditions are summarised. Note that in the inner zone
 the value for $R=R_0$ is assumed as constant in accordance to the
 leading order equation \cref{o1} in which we obtain the conservative
 wave-pinning model.
\begin{table}
\begin{tabular}{|c|c|cc|c|}\hline
\textbf{Zone}  & \textbf{Interval} & \textbf{Boundary Conditions} & & \textbf{regime}\\ \hline
I & $x\in ]-L,x_-]$ & $\frac{d}{dx}\left. (R,S)\right|_{x=-L}=(0,0),$
& $\left. (R,S) \right|_{x=x_-}=(R_0,S_-,)$ & \emph{outer} \\\hline
II & $x\in [x_-,x_+]$ & $\left.(R,S) \right|_{x_-}=(R_0,S_-),$ & $\left.(R,S)\right|_{x_+}=(R_0,S_+)$
 &\emph{inner} \\\hline
III & $x\in [x_+,0]$ &
$\left.(R,S)\right|_{x=x_+}=(R_0,S_+),$ & $ \frac{d}{dx}\left.(R,S)\right|_{x=0}=(0,0)$
 & \emph{outer} \\\hline
\end{tabular}
\caption{ Boundary conditions between the inner and outer regions. The
  interval for each region is specified using the original scale.}
\label{tab:bc}
\end{table}

For the  outer regions we rewrite \cref{sysas} using the scaling \cref{eq:scaling},
the equations at each order in $\varepsilon$ are as follows
\begin{description}
\item[$O(1):$]
\begin{equation}
\label{o1}
F\left( \frac{R+S}{2\delta},\frac{R-S}{2}\right)= O(\eps)=0,
\end{equation}
\item[$O(\varepsilon):$]
\begin{subequations}
\label{orderep}
\begin{align}
\label{orderpR}
\frac{d^2 R}{dX^2}&=\frac{\theta}{2\delta} (R+S)-\alpha,\\
\label{orderpS}
\frac{d^2 S}{dX^2}&=\frac{\theta}{2\delta} (R+S)+\alpha-2\beta_1 \psi+\chi(R,S).
\end{align}

\end{subequations}
\end{description}

The separation of scale at $O(1)$ says that $F$ is approximately zero,
but there is an as yet unspecified  contribution $O(\eps)$. This contribution, termed $\chi(R,S)$,
must be included at next order. In general $\chi$ depends on $R$
and $S$ and it can be seen as a Lagrange multiplier which makes the set of equations
\cref{o1,orderep} a well-defined system of three
differential-algebraic equations for the unknowns $R(X),S(X)$ and
$\chi(R,S)$. \\

 The boundary conditions \cref{tab:bc} ensure the continuity of the solution, but
 they do not provide any information about the parameters involved in
 the asymptotics (namely, $R_0, S_{\pm}, x_{\pm}$). More information
 can be extracted by imposing additional \emph{matching conditions}. A natural
 matching condition for this problem is to impose continuity of the derivatives
\begin{equation}
\left.\frac{d}{dX}(R,S)\right|_{x=x_-}=\left.\frac{d}{dX}(R,S)\right|_{x=x_+},
\label{eq:match}
\end{equation}
which can be sen as being necessary in the limit $\varepsilon \to 0$
in order to satisfy the appropriate jump condition across the inner
zone that are consistent with the second derivative operator.

In summary, the transition under study is equivalent to the
\emph{interior layer} problem, which is given by: the boundary value problems of
the inner region \cref{sysas} and outer regions (\cref{o1,orderep}),
the boundary conditions (specified in \cref{tab:bc})
and the matching condition \cref{eq:match}.\\

In the remainder of this section, by making additional approximations
to this problem, we will show how an analytical close-form
approximate solution can be obtained. We will reduce the
number of unknown parameters to just $R_0$.

 First, since in the inner zone II the dynamics is driven by the
 leading order, we will consider this boundary value problem as
 equivalent to the conservative case and therefore the solution
 \cref{frentecub} is an approximate analytical solution. The problem
 is then reduce to two outer problems (equations \cref{o1,orderep}) for zones I and III,
 whose solutions are given by $(R_1,S_1)$ and $(R_3,S_3)$
 respectively. Under this assumption and considering the boundary
 conditions, the solution in $S$ presents a discontinuity at the point $X=X^*$,
 which connects both outer zones. This assumption is equivalent to saying
 $$X_-=X_+=X^*.$$

When $F$ is given by \cref{dimlecz}, we can write the condition
\cref{o1} in terms of the original variables $(u,v)$ and obtain an
analytical expression for $v(u)$. Expanding it up to linear order
around a certain point $\hat u$, we obtain
\begin{equation}
\label{eq:vdeu}
v(u)=\frac{\eta u(1+u^2)}{1+u^2(1+\gamma)}\approx v(\hat u)+v'(\hat u)(u-\hat u).
\end{equation}
Replacing $u(R,S),v(R,S)$ by:
$$u=\frac{R+S}{2\delta}, \quad v=\frac{R-S}{2},$$ 
and substituting into \cref{eq:vdeu}, we can solve for $S(R)$ and then replace it into \cref{orderpR}, obtaining a second-order affine differential equation. 
The expressions for $S(R)$ and $R(X)$ thus obtained can be written 
\begin{subequations}
\begin{align}
\label{cv}
S(R) &= \left( \frac{\delta -v'(\hat u)}{\delta +v'(\hat u)}\right) R+\frac{2\delta(\hat u v'(\hat u)-v(\hat u))}{\delta+v'(\hat u)},\\
\label{solr}
R(X)&=Ae^{\sqrt{\frac{\theta}{\delta+v'(\hat u)}}X}+Be^{-\sqrt{\frac{\theta}{\delta+v'(\hat u)}}X}\\
&+v(\hat u)+\frac{\alpha \delta}{\theta}+v'(\hat u)\left(\frac{\alpha}{\theta}-\hat u \right),\nonumber
\end{align}
\end{subequations}
where $A$ and $B$ are constants to be determined using the boundary
conditions.  Choosing for Zones I and III
$\hat u=u_0=\alpha/\theta$ and $\hat u= u_+=(R_0+S_+)/(2\delta)$ respectively 
and using the aforementioned boundary conditions, an analytic
approximation for the domain composed by the two outer problems is
\begin{subequations}
\label{eq:sr}
\begin{equation}
\label{eq:srr}
R(X)=\begin{cases}
R_1(X)=\zeta(u_0)+\left(R_0-\zeta(u_0) \right)
\frac{\cosh(\sigma(u_0)(X+\mathcal{L}))}{\cosh(\sigma(u_0)(X^*+\mathcal{L}))}
& X\in[-\mathcal{L},X^*],\\
&\\
R_3(X)=\zeta(u_+)+(R_0-\zeta(u_+))\frac{\cosh(\sigma(u_+)x)}{\cosh(\sigma(u_+)X^*)} & X\in[X^*,0]
\end{cases}
\end{equation}
and
\begin{equation}
\label{eq:srs}
S(X) =\begin{cases} \varphi_1(u_0) R_1(X)+\varphi_0(u_0) & X\in
  [-\mathcal{L},X^*],\\
&\\
\varphi_1(u_+) R_3(X)+\varphi_0(u_+) & X\in [X^*,0].
\end{cases}
\end{equation}
\end{subequations}
Where
\begin{align*}
\zeta(u)&=v(u)+\frac{\alpha  \delta}{\theta}+v'(u)\left(\frac{\alpha}{\theta}-u\right), &\sigma(u) &= \sqrt{\frac{\theta}{\delta+v'(u)}},\\
\varphi_1(u)&= \frac{\delta -v'(u)}{\delta+v'(u)}, & \varphi_0(u)&=\frac{2\delta(uv'(u)-v(u))}{\delta+v'(u)}.
\end{align*}
This solution depends on the parameters
$\{R_0,S_{\pm},X^*\}$. Nevertheless, given $R_0$ we can find
$S_{\pm}$ through \cref{o1}. Moreover, when the dependence between $R$
and $S$ is linear \cref{cv}, the matching condition \cref{eq:match} is
reduced to just the $R$ component
$\left.R_1'(X)\right|_{X^*}=\left.R_3'(X)\right|_{X^*}$, replacing
these expressions from \cref{eq:srr} we obtain:
$$
\frac{(R_0-\zeta(u_0))\sigma(u_0)}{(r_0-\zeta(u_+))\sigma(u_+)}=\frac{\tanh( \sigma(u_+) X^*)}{\tanh(\sigma(u_0)(X^*+\mathcal{L})}.
$$
Additionally, approximating $\tanh(\sigma(u_0)(X*+\mathcal{L}))\approx 1$ and $\tanh(\sigma(u_+) X^*)\approx\sigma(u_+) X^*$, the condition is reduced to:
\begin{equation}
\label{eq:conder}
X^*=\frac{\sigma(u_0)(R_0-\zeta(u_0))}{\sigma(u_+)^2(R_0-\zeta(u_+))}.
\end{equation}

\begin{figure}
\begin{center}
\includegraphics[width=\textwidth]{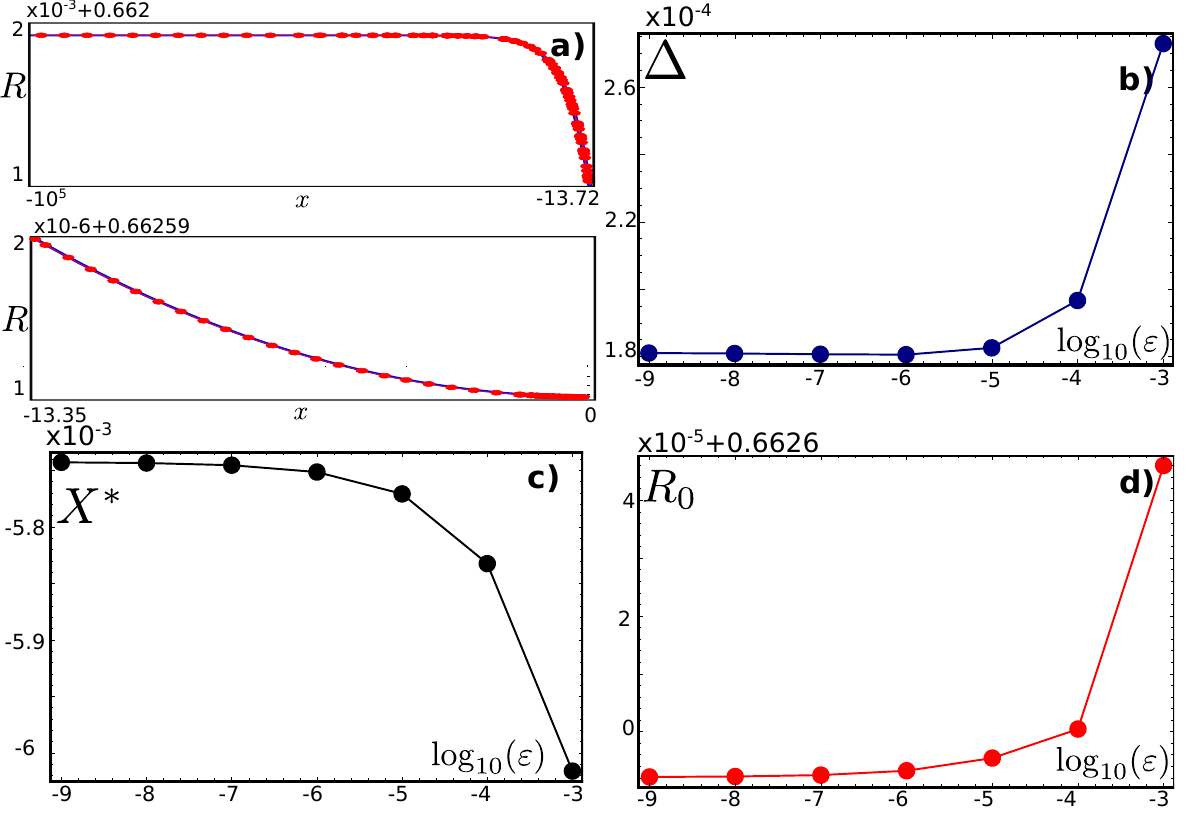}
\end{center}
\caption{Summarising the asymptotic approximation in the case of a
given $R_0$. (a) comparison between the analytical (continuous line) and the numerical solution (red points) of $R$ when $\varepsilon=10^{-9}$, upper and lower 
plots correspond to Zones I and III respectively. (b) 
Difference between the analytical approximation (given by \cref{eq:conder,eq:sr}) and numerical solutions as a function 
of $\varepsilon$. (c) Numerical value of the position of $X^*$ 
as a function of $\varepsilon$. Numerical value of $R$ in Zone II as a function 
of $\varepsilon$. The values of the parameters used are $\delta=2\times 10^{-3},\gamma=14,\eta=5.2,\theta=6,\alpha=1.13$. }
\label{fig:puntos}
\end{figure}

The set of equations \cref{eq:conder,eq:sr} is an analytic
approximation which gives a good approximation to the numerical
solution  when $R_0$ is provided (see \cref{fig:puntos}).
 However, as a consequence of assuming a linear dependence between $R$
 and $S$, the matching conditions for each component \cref{eq:match}
 becomes linearly dependent and therefore the matching condition for $S$ is satisfied
trivially. Therefore, we are unable to determine $R_0$ using this
method. We conjecture that a more accurate matching condition between the Zones I and III is required in order to uniquely determine $R_0$. 

\begin{figure}
\begin{center}
\includegraphics[width=\textwidth]{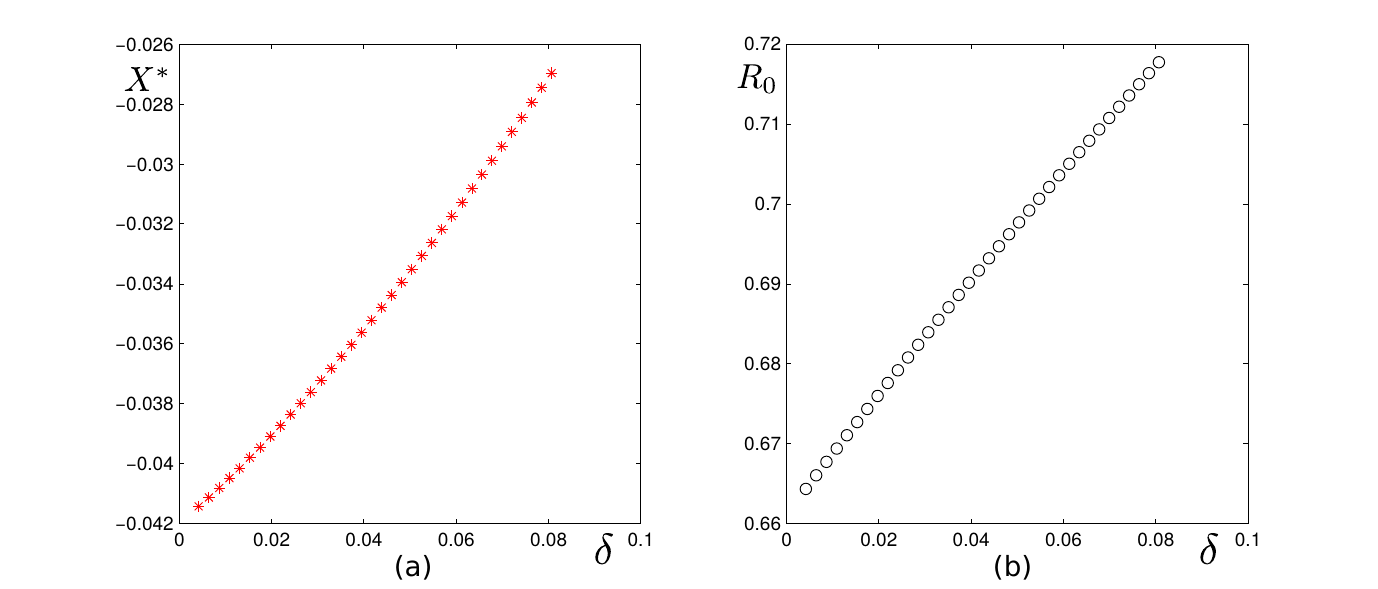}
\end{center}
\caption{Numerical determination of the asymptotic key quantities
  $X^*$ (a) and $R_0$ (b) in \cref{sysas}  as a  function of $\delta$, when
  $\varepsilon=10^{-5},\gamma=14,\eta=5.2,\theta=5.5,\alpha=1.5,L=10^3$. }
\label{fig:delr0}
\end{figure}

In what follows then, we have resorted to numerics to find $R_0$.
Numerically, $R_0$ corresponds to the value of $R$ at $X^*$, which
corresponds to the minimum of $R'$. Using this method, in \cref{fig:delr0} we depicted
 the determined values of $R_0$ and
$X_*$ for a range of values of $\delta$ when the system is in the
quasiconservative regime ($\eps\ll 1$). The rest of the parameters have
been specified in the caption.  From this figure we can appreciate how
the values for $R_0, X^*$  vary continuously with the parameter and therefore they
are well-defined.

   In the \cref{fig:puntos} (a), we have plotted the numerical solution (points) superimposed on the analytical (line)
   when $R_0$ is provided. The solution is illustrated for Zones I and
   III, for the $R$ component. The same comparison for $S$ can be
   obtained from $R$ through \cref{cv}.

In \cref{fig:puntos}(b), the difference between the numerical and the analytical solutions is depicted as a function of $\varepsilon$ . The value of $\Delta$ is given by:
\begin{equation*}
  \Delta=\int_{-L}^{x^-}|R_1(x)-R_n(x)|+|S_1(x)-S_n(x)|dx+\int_{x_+}^0 |R_3(x)-R_n(x)|+|S_3(x)-S_n(x)|dx,
\end{equation*}
where the subscripts $1,3,n$ stand for Zones I, III and the numerical
solutions respectively. Moreover, as the limit is reached, the values
of $R_0$ and $X^*$ (the midpoint between $x_+$ and $x_-$ in the $\varepsilon$-independent scale) attain quickly an almost constant value as $\varepsilon$ tends to zero (cf. \cref{fig:puntos} (c) and (d).

\begin{figure}
\begin{center}
\includegraphics[width=\textwidth]{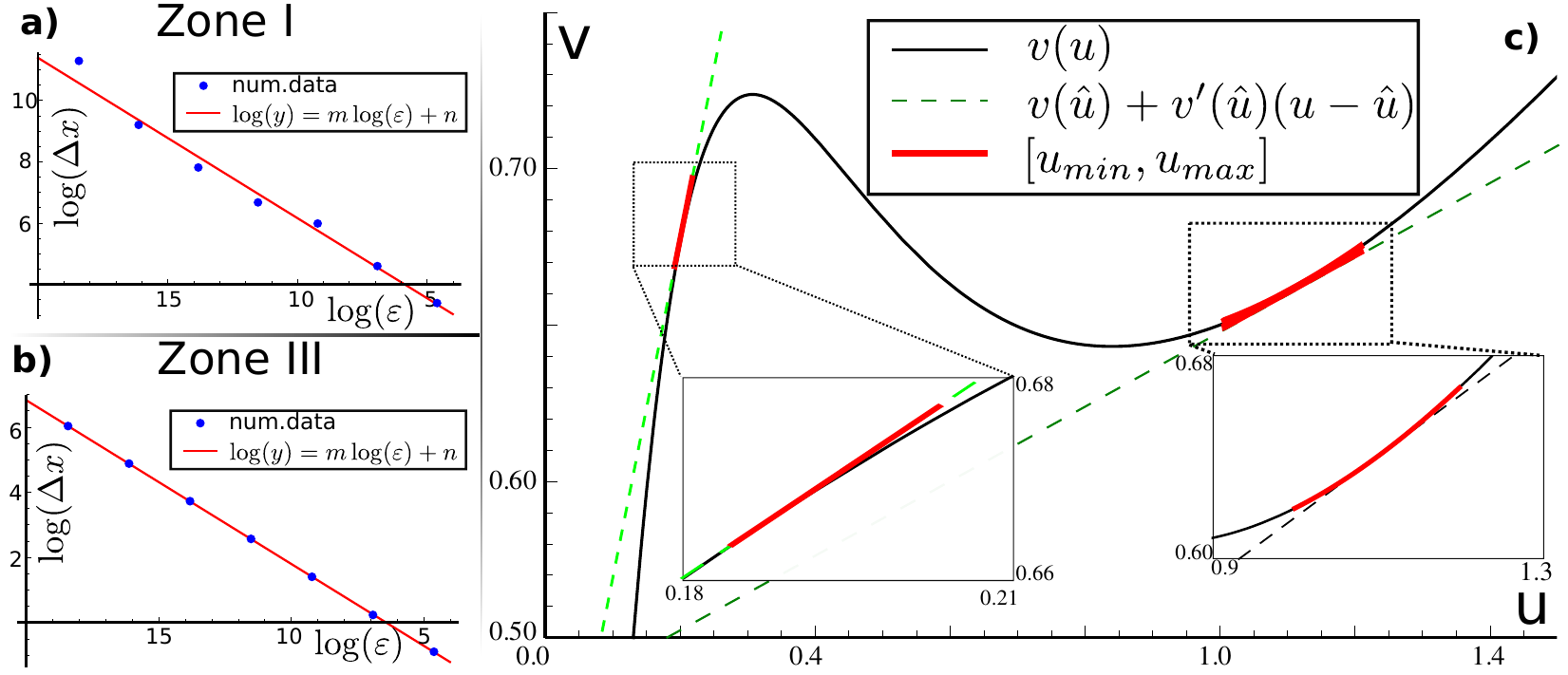}
\end{center}
\caption{(a) Size of Zone I as a function for
  $\varepsilon$ (blue dots), the continuous line correspond to the
  linear fit: $m=-0.522$, $n=0.934$. (b) Size of Zone III as a
  function for $\varepsilon$ (blue dots); the continuous line
  correspond to the linear fit: $m=-0.502$, $n=-3.225$. (c)
  Superposition of $v(u)$ using \cref{eq:vdeu} and the linear
  approximation around $\hat u$ (green dashed line). The range where
  $u$ varies throughout the zone is highlighted in thick red. The
  approximation on the left (right) corresponds to $\hat u=u_0$ ($\hat
  u=u_+$). The values of the parameters used are $\delta = 2\times10^{-3},
  \gamma= 14, \eta = 5.2, \varepsilon= 10^{-9}, \theta= 6, \alpha=
  1.138$.}
 \label{fig:evscal}
\end{figure}

Given $x^*=X^*/\sqrt{\varepsilon}$, we can find numerically $x^{\pm}$
whether using \cref{anchofrente} or by defining $x^+$ ($x^-$) as the
first point to the right (to the left) of $x^*$, where the numerical solution
$(R,S)$ vanishes when is replaced into \cref{o1}. From the numerical values of $x_{\pm}$ it is possible to measure the
length of Zones I and III for several values of $\varepsilon$. In
\cref{fig:evscal} (a) and (b), the $\log -\log$ plot
of the length as function of $\varepsilon$ is presented for both
zones.  The points show a linear behaviour whose slope is approximate
$-1/2$, which is in good agreement with our assumption for the
scaling \cref{eq:scaling}. Moreover, \cref{fig:evscal}(c) depicts a comparison
between the function $v(u)$ (black continuous line) and the linear
approximation (dashed green line) around the two values of $\hat
u=u_0,u_+$.  This figure illustrates how reliable our linear
approximation is to
$v(u)$ in Zones I and III.

\subsection{Finite domain effects}

So far we have been dealing with a domain that is 
large enough for all $\eps$-values considered,
that the localised solution can reach the homogeneous equilibrium at the left-hand
boundary (essentially, an infinite domain). In the conservative case 
(cf.~\cref{sec:wpm} and references therein) the front selection is a property of
the total mass, which itself is greatly affected
by the length of the domain. Hence, a description of the transition to the finite-length case is highly relevant.

Taking the limit $\varepsilon \to 0$ in \cref{eq:sr}, we obtain the
conservative solution \cref{sol:cons}. In contrast, the
asymptotic analysis for a long domain showed a slowly varying behaviour
in the outer zones, with the amplitude (but not the rate) of the
variation being essentially independent the value of $\varepsilon$.
This analysis implicitly assumes that 
Zones I and III can grow to become infinitely long, under the scaling
\cref{eq:scaling} as $\eps \to 0$. Therefore, for a small enough value of
$\mathcal{L}=\sqrt{\varepsilon}L$, size effects will destroy this
picture and the description provided by \cref{eq:sr} will no longer apply. 

The analytical solution \cref{eq:sr} relies on the existence
of a front solution for $S$ which connects $S_{\pm}$ at $x_{\pm}$ while
$R=R_0$ remains constant (Zone II). Beyond that point (Zones I and
III), the solution presents a slow exponential-like behaviour whose
exponents are proportional to $\sqrt{\varepsilon}$. Finally, at the
borders, the derivatives of the solution must vanish. It must be
underlined that our approximate solution \cref{eq:sr} does not have any link with the
homogeneous value at $x=-L$. 

Thus, whenever size effects are
important, the solution at the left border assumes a different value
from the homogeneous equilibrium and eventually reaches $S=S_-$ and
$R=R_0$ as $\varepsilon \to0$. Zone III will increases its size
until it is affected by the finite size effect of Zone I.  At $x=0$, the
value of the solution diminishes (increases) in the $S$ ($R$)
component until it reaches $S_+$ ($R_0$). Through this process, the
limit solution \cref{eq:sr} becomes precisely the front solution without the
slowly varying inner core or tail. As a final
remark, this description explains how the homogeneous value at $x=-L$ in the case
$\eps =0$ can 
depend on the length of the domain.

\begin{figure}
\begin{center}
\includegraphics[width=\textwidth]{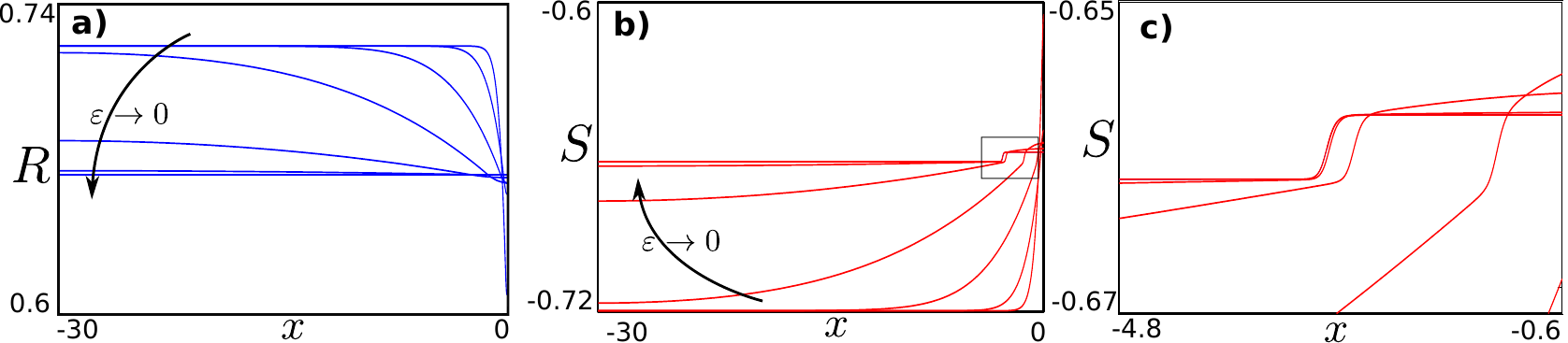}
\end{center}
\caption{Effect of domain size on the homoclinic to heteroclinic transition
as $\eps \to 0$, showing the (a) the $R$-component and (b) the $S$ component, 
in the latter case illustrating the inner zone where the front develops. (c)
A zoom of (b). 
The values of the parameters used are $\delta=2 \times 10^{-3}, \gamma=14,\eta=5.2,\alpha=1.5, \theta=5.5,$ $ \varepsilon=\{0\} \cup \{10^{-n}\}_{n=0}^{10}.$ }
\label{fig:contlfin}
\end{figure}

In the \cref{fig:contlfin}, the above described transition is
illustrated for the finite value $L=30$. When
$\varepsilon=1$ (highest and smallest values of $R$ and $S$
respectively in \cref{fig:contlfin}(a) and (b)),
the solution corresponds to a homoclinic orbit (depicted in half of
the domain). As $\varepsilon$ is diminished, the exponential-like
behaviour in Zone I decreases its exponent according to
\cref{eq:sr}. On the other hand, due to the size restriction and the
boundary conditions, the values of the components change as
$\left.(R(x),S(x)\right)_{x=-L}\to (R_0,S_-)$. Moreover, for a certain
value of $\varepsilon$ (in this case $\varepsilon\approx 10^{-3}$) the
$S-$component becomes a pure front (see the black frame region in
\cref{fig:contlfin}(b)). 
When the front is established,
Zone III can be distinguished and further decreasing of
$\varepsilon$ will displace the front to the left, increasing the size
of Zone III while $\left.(R(x),S(x)\right)_{x=0}\to
(R_0,S_+)$. Eventually the front stops its drift, due to the
conservation of the total mass (\cref{eqc}). In \cref{fig:contlfin}, 
we can observe how the solutions converge to a front for the $S$
component and a homogeneous solution for $R$. The last solution
corresponds to $\varepsilon=0$.

\section{Conclusion}
\label{sec:concl}

The results of this investigation can be collected into 
three main conclusions

First, we studied the \textsl{wave pinning} model, a popular
proposed mechanism for cell polarisation. By introducing the new 
$(R,S)$ coordinates we have added a simpler explanation of how the wave-pinning
works from a mathematical point of view.  Specifically we can then see that
the four-dimensional spatial system is essentially degenerate 
in that it can be represented as a two-dimensional systems with a free parameter
$R_*$. This enables a simple mechanical analogy, the identification of a Maxwell
pinning region and an expression for the front solution in closed form up
to quadrature for any local kinetic function $F$ that has a bistable character.
In certain special cases, an analytic function for the front can be obtained.

Second, motivated by biological systems where there is production and recycling of
G-proteins we have studied the effects of when the mass conservation present in the
original wave-pinning model is relaxed.  The addition of generic \emph{source} and
\emph{loss} terms give rise to several equilibrium solutions such as homogeneous equilibria, periodic patterns, 
localised patterns, and isolated spikes. Moreover, through a combination of
linear stability analysis and numerical continuation we have been able to delineate
how these states are organised in a two-parameter diagram as the diffusion ratio
and nonlinear driving parameter are varied, which relies on the theory
of so-called homoclinic snaking. 
Within this analysis we identified the importance
of a codimension-two bifurcation at which the pattern formation (or Turing) bifurcation
changes from sub- to super-critical and the existence of what we have termed a
Belyakov-Devaney transition in which the pattern precursor is lost.

Finally, we have studied the crucial question of how these two very distinct kinds
of behaviour relate
to each other by taking the limit  $\eps \to 0$, in which
the source and loss terms disappear.  We have shown how the so-called
{\em localisation region} becomes filled with spike solutions and the localised
pattern states disappear. Moreover we have highlighted the delicate asymptotics of
how these spike solutions morph into wave-pinned fronts. In particular this has
involved multiple-scale asymptotic analysis, in which there are three zones,
two outer zones and an inner one. It is only within the inner zone that the front
solution appears. The behaviour in the outer zones is more subtle and depends crucially
on the domain size. 

It might also be interesting to make some more general remarks about
the implication of these results.
Although the model under study  is inspired by the
phenomenon of cell polarisation, the mathematical findings in this
manuscript might be relevant beyond this context. The model
\cref{dimle} is a reaction-diffusion  system which accounts for a
broad spectrum of phenomena. On the other hand, similar models to
\cref{dimle} have been proposed in a rather different context such as
Rho proteins of plants (ROP) (for example \cite{Payne,Victor1}). We
think that the general analysis performed could shed light on the
particular contexts where models similar to \cref{dimle} are used to
describe the dynamics.

 The analysis we have presented, intended to 
be general and is 
independent of the specific form of the local kinetics embodied in
the function $F$.
We have simply used the specific expression \cref{dimlecz} in order to
illustrate the analysis.  More generally,   
a particular result we have found is the 
way that the BD transition causes a non-local bifurcation 
between localised pattern solutions and spike solutions. In effect, the homoclinic snake
is annihilated by the BD transition through the additional peaks
disappearing to the edge of the domain. Given that both
types of localised structures exist in several context, this
transition must be somehow universal. Actually, through private
communications, we have learnt that this transition takes place in
models that have similar properties to \cref{dimle} that arise in:
ecological systems \cite{yuval}, optical systems \cite{lenderkn} and
in a crime wave model \cite{lloyd}. A complete description of the non-local bifurcation and how
it organises localised pattern to spike transition will form the subject of future work. 

There also remain open questions regarding the asymptotic analysis of
\cref{ssec:asym}.  In that study, we fixed all the parameters except $\eps$. In
particular we fixed $\delta \sim 10^{-3}$. Moreover, a necessary
ingredient for  observing non-homogeneous solutions is to consider a small value
of $\delta$ ( see \cite{Mori}, where the asymptotic treatment is
performed in the limit $\delta \ll 1$). Actually, when we studied the
conservative case  in \cref{sec:energy}, we found an analytical approximation for the front
\cref{frentecub}, whose width $\Delta$ (cf. \cref{anchofrente})
depends on $\delta$. Therefore, the asymptotic analysis for the
transition of \cref{ssec:asym}, must be carried out considering  both
$\delta,\eps\ll 1$. This is the next natural step in a better
characterisation of this phenomenon, which may lead to a combination of
singular perturbation theory with the multiple scales asymptotics
studied here
It should also be pointed
out that we have not dealt with the stability of all of these solutions in
the full PDE system in a systematic way. 
Instead we have relied on the persistence of these
solutions observed in the numerical simulations of the PDE. An
analytical treatment of the interaction of the fronts involved in the
heteroclinic loop will be part of future work.

As a final conclusion, it is interesting to note that our analysis suggests
that there is no real distinction between the Turing mechanism, 
wave-pinning or homoclinic snaking
as pattern formation mechanisms in systems of reaction diffusion equations.  
In fully parametrised models, each can be seen as different explanations
that are valid in a different distinguished limits. In a sense, they are
like three sides of the same coin.

\section*{Acknowledgements}
The authors acknowledge helpful conversations with Veronica Greniensen, Stan
Maree, John King, Michael Ward, Lendert Gelens, Yuval Zelnik, Marcel
G. Clerc and Rutuja Patwardhan.
N. Verschueren would like to acknowledge  
``Programa de doctorado en el Extranjero Becas Chile Contract No.72130186.''

\bibliographystyle{plain}
\bibliography{biblio}
\newpage
\appendix

\section{Normal form calculation}
\label{ap:nf}
The goal of this appendix is to determine analytically the co-dimension-two point in
the parameter space where the amplitude of the patterns bifurcates from being
\emph{sub-critical} to \emph{super-critical} in the vicinity of the
spatial instability in the model \cref{eq:dimleap}. In order to find this point, we will compute the amplitude equation for the
patterns up to third order by means of a normal form procedure (see
\cite{ioosbook} for more details). The co-dimension-two point
corresponds to the point, in the spatial instability, where
additionally the third order coefficient in the amplitude equation
vanishes (see \cite{Victor2}). Our starting point is the model
\begin{subequations}
\label{eq:dimleap}
\begin{align}
\partial_t u &=\delta\partial_{xx}u +[F(u,v)-\varepsilon \theta u],\\
\partial_t v &=\partial_{xx}v - [F(u,v)-\varepsilon \alpha]\qquad x\in 
\left[ -L,L \right], \quad \partial_x (u,v) 
(\pm L)=0 , \\
\text{where} & \quad F(u,v) =\left(\gamma \frac{u^2}{1+u^2}+1\right)
v-\eta u= \psi(u) v-\eta u.
\end{align}
\end{subequations}
For sake the of completeness, let us recall the expressions obtained
in \cref{sec:calcula}. The homogeneous equilibrium and the critical curve where the spatial
instability takes place are
\begin{subequations}
\begin{align}
u_0 &= \frac{\alpha}{\theta}, \quad v_0 = \beta_0+\eps \beta_1 =
\frac{\eta u_0}{\psi(u_0)}+\varepsilon \frac{\alpha}{\psi(u_0)},\\
\label{eq:cra}
0&=\eps \theta\psi(u_0)-\frac{(\psi'(u_0)v_0
  -(\psi(u_0)\delta+\eta+\eps\theta))^2}{4\delta},
\end{align}
when:
\begin{equation*}
k_c^2=\frac{\psi'(u_0)v_0-(\psi(u_0)\delta+\eta+\eps \theta)}{2\delta}>0.
\end{equation*}
\end{subequations}
The expression \cref{eq:cra} corresponds to \cref{criticas} written
in terms  of $\psi$ instead of the partial derivatives $\partial_u
F, \partial_v F$ . Evaluating the system \cref{eq:dimleap} at the
critical point (i.e. the
parameters satisfying  \cref{eq:cra}) and translating the system through the change of variables
$$(u,v)=(u_0,v_0)+(U,V),$$
we obtain the main equation of this appendix 
\begin{equation}
\label{eq:equnia}
\partial_t \left(\begin{array}{c} U\\ V \end{array} \right)=
[\mathbb{J}+\mathbb{D}\partial_{xx}]_{c}\left(\begin{array}{c} U\\ V \end{array} \right)+\left(\begin{array}{c}
1\\ -1\end{array}\right) \mathbb{NL} (U,V),
\end{equation}
where
\begin{align}
\label{eq:operador}
[\mathbb{J}+\mathbb{D}\partial_{xx}]_{c} &= \left[ \begin{array}{cc}
    \delta \partial_{xx} +\psi'(u_0) v_0 -(\eta+\eps \theta) &
    \psi(u_0) \\
\eta-\psi'(u_0)v_0 & \partial_{xx} -\psi(u_0) \end{array} \right]_{c},\\
\label{eq:nonl}
\mathbb{NL} (U,V)&=\psi'(u_0)U V+(v_0+V)\sum_{n=2}^{\infty}
\frac{\psi^{(n)}(u_0)}{n!} U^{n}. 
\end{align}
Here, the sub-index $c$ in the linear operator stands for the critical
values of the parameters when one of them is fixed through
\cref{eq:cra}. In the remainder we will drop this sub-index assuming
that we are at the critical point. Using \cref{eq:operador,eq:nonl}, we will compute the
amplitude equation by using the normal form procedure. More precisely, we will look for the
amplitude equation \cref{eq:fnansaa}, and the change of variables \cref{eq:fnanscva} (which transforms from
the original variables $(u,v)$ into the new variable $A$) at the same
time. This is respectively
\begin{subequations}
\label{eq:fnansap}
\begin{align}
\label{eq:fnanscva}
\left( \begin{array}{c} U\\ V\end{array}\right) &=W^{[1]}+W^{[2]}+\ldots,\\
\label{eq:fnansaa}
\partial_t A&=\partial_t A^{[1]}+\partial_t A^{[2]}+\ldots,
\end{align}
\end{subequations}
where the superscript accounts for the order in $A$ (in the remainder
we will refer to this as ``the order''). Formally, this equation must
be solved at each order until it saturates (until the first coefficient obtained is negative in the region of
parameters of interest).  At that point, the amplitude equation is
obtained \emph{at the critical point}. An extra step, the \emph{unfolding}, is necessary in
order to allow the critical parameter to present a small variation
around its critical value. Since we are only interested in the change of
sign in the third coefficient, we will limit our calculations to that order.
We are now in conditions to start the \emph{normal form} procedure

\begin{flushleft}
\textbf{Order 1}
\end{flushleft}

At order 1, \cref{eq:equnia} is
\begin{equation*}
\partial_A W^{[1]} \partial_t A^{[1]}+c.c.=
[\mathbb{J}+\mathbb{D}\partial_{xx}] W^{[1]},
\end{equation*}
here $c.c.$ stands for \emph{complex conjugate}. This abbreviation
will be used in the rest of the appendix. Since we are interested in constructing a correction to the pattern
solution, the first order corresponds to the linear approximation. This
is equivalent to consider $\partial_t A^{[1]}=0$. Replacing the
following ansatz
$$W^{[1]}=\left( \begin{array}{c} w_1 \\ w_2 \end{array}\right) (Ae^{i
  k_c x}+\bar{A} e^{-i k_c x}),
$$
and keeping in mind that the linear operator at the critical point is
\emph{singular}, the solution for $W^{[1]}$ is
\begin{equation}
W^{[1]} =\left( \begin{array}{c}
\psi(u_0)\\
   \delta k_c^2-\psi'(u_0)v_0 +\eta+\varepsilon \theta\end{array}\right)(Ae^{i  k_c x}+c.c.).
\label{eq:kerna}
\end{equation}
This expression is the same as \cref{eq:linsol}. 
\begin{flushright}
\textbf{End of Order 1.}
\end{flushright}

\begin{flushleft}
\textbf{Order 2}
\end{flushleft}

At order 2, \cref{eq:equnia} is
\begin{subequations}
\label{eq:orden2}
\begin{align}
\label{eq:or2a}
\partial_A W^{[1]} \partial_t A^{[2]}+c.c. &=
[\mathbb{J}+\mathbb{D}\partial_{xx}] W^{[2]}+\left(\begin{array}{c}
1\\ -1\end{array}\right)( (|A|^2+ c.c.)b_{20}+(A^2e^{2ik_c
x}+c.c)b_{22}),\\
\label{eq:or2b}
b_{20}&=\frac{1}{2} w_1 \left(v_0 \psi ''(u_0) 
  w_1+2 \psi '(u_0) w_2\right)=b_{22}.
\end{align}
\end{subequations}
Since in the nonlinear terms there are not \emph{secular terms} (terms
which are proportional to the kernel of the linear operator
\cref{eq:kerna}), we can take $\partial_t A^{[2]}=0$. Actually, this is
the case for every \emph{even order} (i.e. $\partial_t A^{[2n]}=0$). Taking into account this assumption, we can solve the equation at this order
  by proposing the following ansatz
\begin{equation}
\label{eq:ansat}
W^{[2]}=\varphi_{20}( |A|^2+c.c.)+\varphi_{22} (A^2 e^{2ik_c
  x}+c.c.).
\end{equation}
In the remainder, we will carry out an analogous procedure for the
different orders. Hence, it is useful to introduce the following
notation to make the steps more clear. At the different orders, the
scalar quantity multiplying different powers of $A$ will be denoted by
$b_{i,j}$ (see for example $b_{20}$ in \cref{eq:or2b}), where the first sub-index stands for the order and the 
second for the multiple of $k_c$ in the exponent. The same
labels are used in our ansatz \cref{eq:ansat} for the vectors
$\varphi_{i,j}$. At this order, the uncoupled  linear systems obtained
are
$$\mathbb{J}\varphi_{20}=b_{20}\left(\begin{array}{c}-1\\
    1\end{array}\right),$$
$$[\mathbb{J}-4
k_c^2\mathbb{D}]\varphi_{22}=b_{22}\left(\begin{array}{c}-1\\1\end{array}\right).$$
Solving this equations we obtain the first correction in the change of
variables \cref{eq:fnanscva}.

\begin{flushright}
\textbf{End of Order 2.}
\end{flushright}

\begin{flushleft}
\textbf{Order 3}
\end{flushleft}
Using the same notation introduced in the previous order, at order 3
the equation \cref{eq:equnia} is

\begin{align}
\label{eq:o3}
[\mathbb{J}+\mathbb{D}\partial_{xx}] W^{[3]}&=\left(\begin{array}{c}-1\\ 1\end{array}\right) [b_{31} (|A|^2 A e^{ik_c x}+c.c.)+b_{33}(A^3
e^{3ik_c x}+c.c.)]\nonumber\\
& +(\partial_A W^{[1]} \partial_t A^{[3]}+c.c),
\end{align}
with the coefficients
\begin{align*}
b_{31}&= \frac{1}{6} \left\{3 v_0 \psi ^{(3)}(u_0) w_1^3+12 v_0 \psi ''(u_0) w_1 \varphi_{20}^1+6 v_0 \psi ''(u_0)
   w_1 \varphi_{22}^{1}+9 \psi ''(u_0) w_2 w_1^2\right.\\
&\left.+12 \psi '(u_0) w_1   \varphi_{20}^2+12 \psi '(u_0) w_2\varphi_{20}^1+6 \psi '(u_0) w_1
  \varphi_{22}^2+6 \psi '(u_0) w_2 \varphi_{22}^1\right\},\\
b_{33}&=\frac{1}{6}\left(v_0 \psi ^{(3)}(u_0) w_1^3+6 v_0 \psi
  ''(u_0) w_1 \varphi_{22}^1+3 \psi ''(u_0)w_2 w_1^2\right)+ \psi '(u_0)w_1 \varphi_{22}^2+ \psi'(u_0) w_2 \varphi_{22}^1.
\end{align*}

In contrast with the previous case, here we can notice the existence
of secular terms (this is the case for every \emph{odd order}). In
order to ensure the solvability of this equation we make use of the
\emph{solvability condition} (or \emph{Fredholm Alternative}
theorem)\cite{ioosbook}.  Considering the following inner product
\begin{equation}
\langle \vec v_0 f(x) |\vec w_0 g(x)\rangle =
\frac{1}{X}\int_{y}^{y+X} f^*(x) g(x) dx  \vec v_0 \cdot \vec
w_0,
\label{eq:pea}
\end{equation}
the adjoint of the linear operator is:
$$[\mathbb{J}+\mathbb{D}\partial_{xx}]^{\dagger}
=[\mathbb{J}^{t}+\mathbb{D}\partial_{xx}],$$
and therefore the kernel of this operator is:
$$W^{\dagger}=\left(\begin{array}{c}w_1^{\dagger}\\
w_2^{\dagger} \end{array} \right)(e^{i  k_c x}+ e^{-i k_c x})=
\left(\begin{array}{c}\psi'(u_0)v_0-\eta\\
-\delta k_c^2+\psi'(u_0)v_0-\eta-\varepsilon \theta \end{array} \right)(e^{i  k_c x}+ e^{-i k_c x}).
$$
Hence, according to the solvability condition, it is enough to ensure
the orthonormality between the right hand side of \cref{eq:o3} and
the kernel of the adjoint operator to ensure the solvability of
problem \cref{eq:o3}.

 Moreover, under the scalar product given by
\cref{eq:pea}, the  following \emph{orthonormality} relation is
satisfied
$$\frac{1}{X}\int_{y}^{y+X} e^{-im k_c x} e^{in k_c x} dx =\delta_{n,m}.$$
Then, the \emph{solvability condition} is reduced in this case to the
following expression
$$ \partial_t A^{[3]} 
\langle\left(\begin{array}{c} w_1 \\ w_2 \end{array}\right) |
\left(\begin{array}{c} w_1 ^{\dagger}\\ w_2
    ^{\dagger} \end{array}\right)
 \rangle 
+\langle\left( \begin{array}{c} -1 \\ 1 \end{array}\right) | \left(\begin{array}{c} w_1 ^{\dagger}\\ w_2
    ^{\dagger} \end{array}\right)b_{31}\rangle |A|^2 A =0,$$
solving for $\partial_t A^{[3]}$
$$\partial_t A^{[3]} =\frac{\langle\left( \begin{array}{c} 1 \\ -1 \end{array}\right) | \left(\begin{array}{c} w_1 ^{\dagger}\\ w_2
    ^{\dagger} \end{array}\right)\rangle}{ \langle\left(\begin{array}{c} w_1 \\ w_2 \end{array}\right) |
\left(\begin{array}{c} w_1 ^{\dagger}\\
    w_2^{\dagger} \end{array}\right) \rangle }b_{31} |A|^2 A=C_3 |A|^2 A.$$

\begin{figure}
\begin{center}
\includegraphics[width=12.4cm]{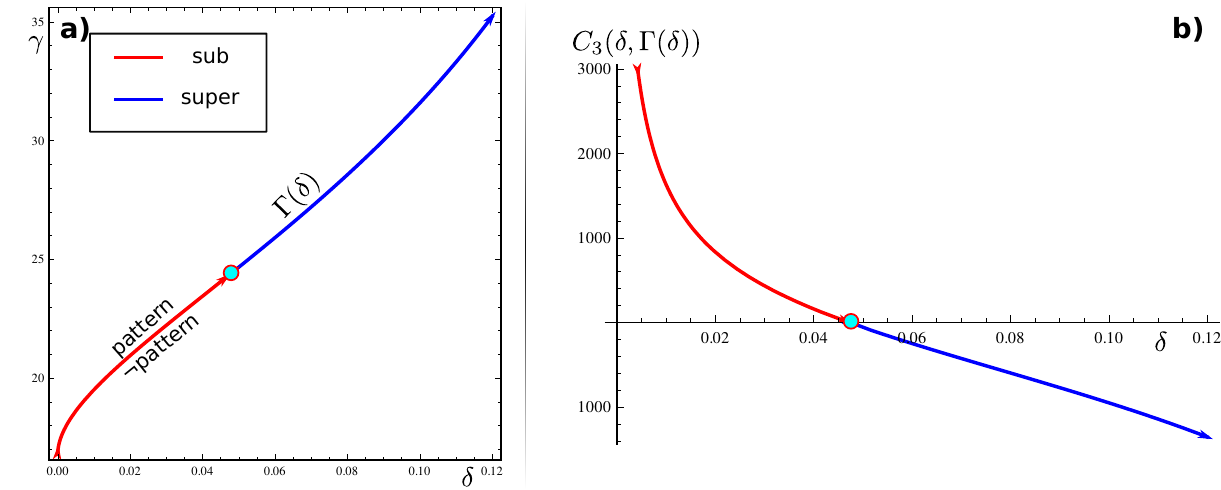}
\end{center}
\caption{ Determination of the co-dimension-two point  for the
  parameter values $\varepsilon=1,\gamma\in
  [11,35],\delta\in[0,0.12],\eta=5.2, \theta=5.5,\alpha=1.5$
  \textbf{a)} The critical curve has been parameterised as
  $\Gamma(\delta)$. In \textbf{b)}, a plot of $C_3$ as a function of
  the critical curve is presented. The existence of a zero (light blue
dot) changes the nature of the amplitude equation from sub-critical
(red) and super-critical (blue).}
 \label{fig:ap}
\end{figure}

By imposing this condition  in \cref{eq:o3}, we can solve the
equation at this order by introducing the ansatz
$$W^{[3]}=\varphi_{31}( |A|^2A e^{ik_c x}+c.c.)+\varphi_{33} (A^3
e^{3ik_c x}+c.c.)$$
The uncoupled \emph{solvable} linear systems are obtained
\begin{align*}
[\mathbb{J}-k_c^2\mathbb{D}]\varphi_{31}=&\left(\begin{array}{c}w_1\\w_2\end{array}\right)
C_3 +\left( \begin{array}{c} -1\\ 1\end{array}\right) b_{31}\\
[\mathbb{J}-9k_c^2\mathbb{D}]\varphi_{33}=&\left( \begin{array}{c} -1\\ 1\end{array}\right) b_{33}
\end{align*}
 By solving this system we obtain the change of variables at this order
\begin{flushright}
\textbf{End of Order 3.}\\
\end{flushright}
\newpage
Now that an analytic expression for cubic coefficient $C_3$ has been determined, we can
look for a change of sign along the critical curve.\\
 All the corrections and coefficients obtained throughout this procedure depend on
the previous orders. Even though obtaining closed-form  expressions is
possible, it is in general cumbersome. Hence, the use of a symbolic
algebra software is strongly recommended.  In order to illustrate our
calculations, \cref{fig:ap} shows the existence of a
co-dimension-two point in the parameter space. Allowing $\delta$
and $\gamma$ to vary  and fixing the rest of the parameters as in
\cref{sec:calcula}, we parameterise the critical curve
\cref{eq:cra} (blue and red line in \cref{fig:ap} (a)) as a function
$\Gamma(\delta)$). Hence, $C_3(\Gamma(\delta),\delta)$ is a single
variable function which presents one zero (light blue dot in
\cref{fig:ap}), changing the nascence of spatial patterns from being
\emph{super-critical} (blue) to \emph{sub-critical} (red). The existence
of a region where a sub-critical amplitude take place guarantee the
existence of a bi-stability region between homogeneous equilibrium and
spatial patterns. Within that region, a sub-region where localised
structures exist could be found. Therefore, a point in the parameter space below the critical
curve and in the sub-critical region is an educated guess of a
starting point in the search
for localised structures.  As a final remark, when the sub-critical
bifurcation for the amplitude takes place, as we pointed out above, a
higher order correction must stabilise the spatial patterns. From the
numerics on the full model, we know that the patterns are stable and
therefore, this further correction have not been computed.

\end{document}